\documentclass[sts]{imsart}

\RequirePackage{amsthm,amsmath,amsfonts,amssymb}
\RequirePackage[square,numbers]{natbib}
\RequirePackage[colorlinks=false,citecolor=black,urlcolor=black]{hyperref}

\usepackage{tikz}
\usepackage{adjustbox}
\usetikzlibrary{bayesnet}
\usepackage[acronym,smallcaps]{glossaries}
\usepackage{colortbl}
\PassOptionsToPackage{usenames,dvipsnames,svgnames,table}{xcolor}  

\usepackage{bm}

\newcommand\Nr{U} 
\newcommand\nr{u} 
\newcommand\Nt{N} 
\newcommand\nt{n} 
\newcommand\daysub{t}
\newcommand\weeksub{w}
\newcommand\ltlasub{i}
\newcommand\phesub{J}
\newcommand\prevnum{I}
\newcommand\prevprop{\pi}

\newcommand\popsize{M}
\newcommand\subit{{\ltlasub,\daysub}}
\newcommand\subiw{{\ltlasub,\weeksub}}

\newcommand\subjw{{\phesub,\weeksub}}
\newcommand\Rt{\mathcal{R}^{\text{eff}}}
\newcommand\react{\text{REACT}}
\newcommand\ons{\text{ONS CIS}}
\newcommand\pill{\text{Pillar1+2}}
\newcommand\pilln{\nt\text{ of }\Nt}
\newcommand\reactn{\nr\text{ of }\Nr}
\newcommand\pillnjw{\nt_\subjw\text{ of }\Nt_\subjw}
\newcommand\pillniw{\nt_\subiw\text{ of }\Nt_\subiw}
\newcommand\reactnjw{\nr_\subjw\text{ of }\Nr_\subjw}

\newcommand{\Emod}{\mathcal{M}_{\text{E}}}
\newcommand{\Dmod}{\mathcal{M}_{\text{D}}}
\newcommand{\EDmod}{\mathcal{M}_{\text{ED}}}

\newcommand\pcr{\prevnum}
\newcommand\wtpcr{\text{PCR+}}
\newcommand\lfd{\text{\#LFD+}}
\newcommand\infs{\text{\#Infectious}}

\newcommand\imm{\text{\#Immune}}

\newcommand\dtest{D^{\text{Test+}}}
\newcommand\dpcr{D^{\text{PCR+}}}
\newcommand\inc{\text{\#New cases}}
\newcommand\npi{\text{NPIs}}
\newcommand\mobility{\text{Mobility}}
\newcommand\mobilitydata{\text{Mob data}}
\newcommand\serology{\text{Serology}}
\newcommand\vax{\text{Vaccin.}}
\newcommand\data{\text{Other data}}
\newcommand\datapar{\B{\theta}}
\newcommand\variant{\text{Variants}}

\newcommand\lfdtest{\text{LFD}}
\newcommand\ntofN{\nt\text{\ of\ }\Nt}
\newcommand\nrofN{\nr\text{\ of\ }\Nr}

\newcommand\biaspar{\text{Bias}}   
\newcommand\biasparpill{\delta}   
\newcommand\biasparlfd{\biaspar_{\text{LFD}}}   
\newcommand\biasparons{\biaspar_{\text{Household}}}   
\renewcommand\contentsname{}

\newcommand\mdebias{\mathcal{M}_{\text{Debias}}}   
\newcommand\mnpi{\mathcal{M}_{\text{NPI}}}   
\newcommand\mst{\mathcal{M}_{\text{ST}}}   

\newcommand{\B}[1]{\boldsymbol{#1}}

\renewcommand\contentsname{}
\newcommand\Pp{\mathbb{P}}

\newcommand\Tes{\text{Tested}}

\newcommand\Infd{\text{Infected}}

\newcommand\namerefquote[1]{\textit{\nameref{#1}}}

\definecolor{gn}{rgb}{.0, .5, .75}
\definecolor{bl}{rgb}{.5, .25, .75}
\definecolor{ch}{rgb}{.5, 0, .5}
\definecolor{mbr}{rgb}{.5, 1, .5}
\definecolor{sr}{rgb}{0, 0.5, 0}
\definecolor{pd}{rgb}{0.5, 0.5, 0}
\definecolor{tp}{rgb}{1, 0, 0}
\definecolor{mbl}{rgb}{1, 0, 0}
\definecolor{rj}{rgb}{0, 1, 0}
\definecolor{ywt}{rgb}{0, 0, 1}
\definecolor{hg}{rgb}{0, .5, 0}
\definecolor{tef}{rgb}{.5, 0, .5}

\newacronym{phe}{PHE}{Public Health England}
\newacronym{nhs}{NHS}{National Health Service}
\newacronym{sarscov2}{SARS-CoV-2}{severe acute respiratory syndrome coronavirus 2}
\newacronym{uk}{UK}{United Kingdom}
\newacronym{covid}{COVID-19}{coronavirus disease 2019}
\newacronym{pcr}{PCR}{polymerase chain reaction}
\newacronym{npi}{NPI}{non-pharmaceutical intervention}
\newacronym{ons}{ONS}{Office for National Statistics}
\newacronym{cis}{CIS}{COVID-19 Infection Survey}
\newacronym{react}{REACT}{REal-time Assessment of Community Transmission}
\newacronym{dag}{DAG}{directed acyclic graph}
\newacronym{ltla}{LTLA}{lower tier local authority}
\newacronym{eb}{EB}{empirical Bayes}
\newacronym{ci}{CI}{credible interval}
\newacronym{voc}{VoC}{variant of concern}
\newacronym{sgtf}{SGTF}{S gene target failure}
\newacronym{sir}{SIR}{Susceptible-Infectious-Recovered}
\newacronym{dtmc}{DTMC}{discrete time Markov chain}
\newacronym{si}{SI}{Supplementary Information}
\newacronym{bame}{BAME}{Black, Asian and Minority Ethnic}
\newacronym{dd}{DD}{data-dependent}
\newacronym{imd}{IMD}{indices of multiple deprivation}
\newacronym{wc}{w/c}{the week commencing}

\startlocaldefs

\endlocaldefs

\begin{document}

\begin{frontmatter}
\title{Interoperability of statistical models in pandemic preparedness: principles and reality}
\runtitle{Interoperability and pandemic preparedness}

\begin{aug}

\author[1,+]{\fnms{George} \snm{Nicholson}\textsuperscript{1,+}},
\author[2,+]{\fnms{Marta} \snm{Blangiardo\textsuperscript{2,+}}},
\author[9]{\fnms{Mark} \snm{Briers}},
\author[7,+]{\fnms{Peter J} \snm{Diggle\textsuperscript{7,+}}},
\author[9,+]{\fnms{Tor Erlend} \snm{Fjelde\textsuperscript{9,+}}},
\author[9,+]{\fnms{Hong} \snm{Ge\textsuperscript{9,+}}},
\author[8]{\fnms{Robert J B} \snm{Goudie\textsuperscript{8}}},
\author[5,+]{\fnms{Radka} \snm{Jersakova\textsuperscript{5,+}}},
\author[6,+]{\fnms{Ruairidh E} \snm{King\textsuperscript{6,+}}},
\author[1,+]{\fnms{Brieuc C L} \snm{Lehmann\textsuperscript{1,+}}},
\author[6,+]{\fnms{Ann-Marie} \snm{Mallon\textsuperscript{6,+}}},
\author[2,+]{\fnms{Tullia} \snm{Padellini\textsuperscript{2,+}}},
\author[1,+]{\fnms{Yee Whye} \snm{Teh\textsuperscript{1,+}}},
\author[1,5,6,+,*]{\fnms{Chris} \snm{Holmes\textsuperscript{1,5,6,+,}*}} and
\author[5,8,+,*]{\fnms{Sylvia} \snm{Richardson\textsuperscript{5,8,+,}*} \ead[label=e3]{sylvia.richardson@mrc-bsu.cam.ac.uk}}

\address[1]{\textsuperscript{1}University of Oxford, UK.}
\address[2]{\textsuperscript{2}MRC Centre for Environment and Health, Dept of Epidemiology and Biostatistics, Imperial College London.}
\address[3]{\textsuperscript{3}Health Economics Research Centre, Nuffield Department of Population Health, University of Oxford, UK.}
\address[4]{\textsuperscript{4}The National Institute for Health Research Health Protection Research Unit in Healthcare Associated Infections and Antimicrobial Resistance at the University of Oxford, University of Oxford, Oxford, UK.}
\address[5]{\textsuperscript{5}The Alan Turing Institute, London, UK.}
\address[6]{\textsuperscript{6}MRC Harwell Institute, Harwell, UK.}
\address[7]{\textsuperscript{7}CHICAS, Lancaster Medical School, Lancaster University, UK.}
\address[8]{\textsuperscript{8}MRC Biostatistics Unit, University of Cambridge, UK, \printead{e3}}
\address[9]{\textsuperscript{9}Department of Engineering, University of Cambridge, UK.}
\address[*]{*These authors contributed equally to this research.}

\address[+]{\textsuperscript{+}Members of The Alan Turing Institute and Royal Statistical Society's ``Statistical Modelling and Machine Learning Laboratory'', in partnership with the Joint Biosecurity Centre, part of NHS Test and Trace within the Department of Health and Social Care.\url{https://www.turing.ac.uk/research/research-projects/providing-independent-research-leadership-joint-biosecurity-centre}}

\end{aug}

\begin{abstract}
We present ``interoperability'' as a guiding framework for  statistical modelling to assist policy makers asking multiple questions using diverse datasets in the face of an evolving  pandemic response. Interoperability provides an important set of principles for future pandemic preparedness, through the joint design and deployment of adaptable systems of statistical models for disease surveillance using probabilistic reasoning. We illustrate this through case studies for inferring spatial-temporal \gls*{covid} prevalence and reproduction numbers in England.
\end{abstract}

\begin{keyword}
\kwd{Bayesian graphical models}
\kwd{COVID-19}
\kwd{Evidence synthesis}
\kwd{Interoperability}
\kwd{Modularization}
\kwd{Multi-source inference}
\end{keyword}

\end{frontmatter}

\section{Background and key principles of interoperability} 
\label{background}
Faced with the \gls*{sarscov2} pandemic that posed an urgent and overwhelming threat to global population health, policy makers worldwide sought to muster  reactive and proactive analytic capabilities in order to track the evolution of the pandemic in real time, and to investigate potential control strategies. In the UK,  governmental health data analytic capabilities were strengthened in the midst of the \gls*{covid} pandemic with the creation in May 2020 of the Joint Biosecurity Centre (JBC), whose mission is to provide evidence-based, objective analysis, assessment and advice so as to inform the response of local and national decision-making bodies to  current and future epidemics. In parallel, the JBC established links with academic groups including a partnership between The Alan Turing Institute (herein Turing) and the Royal Statistical Society (RSS), leading to the creation in October 2020 of the "Statistical Modelling and Machine Learning Laboratory" (the Lab) within JBC. The Turing-RSS Lab's aims are to work within the JBC to provide additional capacity through independent, open-science research based on rigorous statistical modelling and inference directed at JBC priority areas.
  
Established against the background of a fast moving pandemic, this partnership brought into focus a number of interesting challenges to conventional statistical practice arising, in particular, from the need to model real-time, messy data from diverse sources, in order to efficiently address rapidly evolving public health demands. The dynamic nature of the pandemic and the resulting public health priorities led to frequent changes in the research questions being asked of the data, with focus often shifting unpredictably and suddenly. This challenged conventional statistical analysis protocols that target specific research questions, as these would take too long to deliver, and carry an associated risk of redundancy. Instead, it was necessary to develop robust, easy-to-update and reuseable modules
which could be integrated into arbitrarily complex models to provide analyses useful for decision-making (we will clarify  
 our use of the terminology ``\emph{model}'' and ``\emph{module}'' at the end of this section). 
 
Our practice and strategic thinking led us to develop a set of inter-connected health protection models, and to articulate the principle of ``{\em{interoperability}}'' as an important statistical concept and goal for future disease surveillance systems.\footnote{We stress at the outset that when we refer simply to ``interoperability'', we always mean statistical interoperability. The concept of interoperability referred to in engineering or for computer hardware is not the subject of this paper.} In this article, we discuss the emerging principles of interoperability of statistical models that we  have started to operationalise, and we illustrate these on case studies carried out in the Lab. Interoperability can be usefully, and loosely,  characterised as providing an operational statistical and computational data-driven framework which is designed to optimally fulfil the following complementary goals: 
\begin{itemize} 
    \item \emph{Effective information synthesis.} When faced with multiple and evolving datasets, aim for transparent and appropriate sharing of information across diverse data modules that may be misspecified and/or conflict with one another;
    \item \emph{Synchronising high-quality data across models.} Maintain version-controlled data streams synchronised across a system of models and modules, accompanied by dynamic, context-specific, semi-automated quality control, e.g.\ through global visualization of data inputs and results;
    \item \emph{Coherence of findings.} When building models to answer different questions, harmonize the models' essential elements by incorporating common key latent quantities, such as disease prevalence;
     \item \emph{Inter-connectivity.} Use modular statistical approaches, such as Bayesian graphical models, to ensure principled propagation of uncertainty and to aim for structural robustness where appropriate, e.g. through the concept of \emph{loose coupling} \cite{stevens_structured_1974,offutt_software_1993}, in which the failure/misspecification of one module does not adversely affect other modules;
      \item \emph{Cross-talk between computational tools.} The ability to transfer results of analyses between different computational tools, with the freedom to implement each module of analysis primarily in its most convenient software package;
      \item \emph{Reproducibility.} Ensure the generation of stable reproducible results via an open-source code base of inter-connectable models and plug-in analysis modules, coupled with tight control on data versioning;
    \item \emph{Consistent model validation.} Identify opportunities for ongoing dynamic assessment of model performance, ideally that can be applied to multiple models sharing latent quantities and/or addressing similar questions.
    \end{itemize}
 
 There are obvious connections between these objectives. The key desirability of modularity informed our choice of Bayesian inference methods and Bayesian graphical models as the core statistical framework. We have heavily relied on the known flexibility of Bayesian hierarchical models (BHM) with latent (Gaussian) process components to deliver predictive and/or explanatory inferences whilst accommodating complex data structures and measurement processes and enabling principled data synthesis (for examples, see \cite{ades_multiparameter_2006} and the introductory chapter of \cite{green_highly_2003}). Recent work on Markov melding \cite{goudie_joining_2019} and statistical learning while cutting or restricting information flow between modules  \cite{lunn_combining_2009,bayarri_modularization_2009,plummer_cuts_2015,jacob_better_2017,carmona_semi-modular_2020} is particularly relevant to anchor and operationalize our goals. 

We will often use the terminology ``model'' alongside the distinct but related terminology ``module/modular/modularity'', so it is useful to compare and contrast these concepts at the outset:
\begin{itemize} 
    \item \emph{Model.} ``A statistical model is a probability distribution constructed to enable inferences to be drawn or decisions made from data.'' \cite{davison_statistical_2003}. A model can be comprised of one or more modules. 
    \item \emph{Module.} A module is a model component or, more precisely, a joint probability distribution linking some or all of: observed data, latent (unobserved) parameters or data, and user-specified hyperparameters. We use ``\emph{data module}'' to refer to a data-containing module. In contrast to a model, a module need not be equipped to provide stand-alone inference from data; therefore we introduce the term ``\emph{analysis module}'' to denote an inference-ready module, possibly paired with an analytical method. ``\emph{Modular}'' methods, and the general principle of ``\emph{modularity}'', allow a model to be decomposed into smaller, less complex modules, thereby facilitating one or more of: efficient/flexible computation, analytical simplification, balanced information flow, and/or structural robustness. 
    \end{itemize}
Of course the distinction between model and module is not strict, as a simple model may also be a module of a more complex model. Rather than trying to be pedantic with the above definitions, we are simply aiming to convey the spirit and sense of the vocabulary 
we use throughout the paper.

\section{Introduction to interoperability as a strategic goal, informed by current approaches to disease surveillance}

Integrated infectious disease surveillance has unusual characteristics from a statistical analysis perspective. Multiple research questions on disparate outcomes are targeted towards better understanding {\em{of a common underlying process}}, e.g.\ of the disease spread and its evolution in time and space. Asking multiple questions of a single process opens up a spectrum of modelling approaches. At one extreme, different models using potentially partially overlapping data could be built independently to estimate common latent characteristics of the disease under consideration.
At the other extreme, we can look to build a single universal joint model covering every facet of the disease process that is theoretically able to answer any question. Our conjecture is that, in the face of the constraints and operational challenges outlined above, there is an optimal middle ground in which we develop models and analysis plans and couple them according to the guiding principles of interoperability. 


\subsection{Building separate models for common key latent quantities}

Many flavours of epidemic models have been developed for tracking \gls*{covid} disease transmission ranging from agent-based simulation models to age-structured compartmental models or discretized semi-mechanistic models using renewal equations \cite{anderson_2020_royal_society,flaxman2020estimating}. It is not our purpose to give a comprehensive review of these but simply to take stock of the rich diversity of modelling approaches and data sources chosen by academic groups to inform the calibration or estimation of epidemic parameters. In the UK,  a number of academic modelling groups have been actively participating in the expert advisory panel SPI-M, a subgroup of SAGE, which has advised the UK government from the start of the pandemic. The adopted collegiate mode of working of SPI-M has fostered the development of a set of distinct models for estimating key epidemic quantities and producing short-term forecasts. 
Besides the differences between the modelling approaches, the choice of primary data sources and the ways to embed this information into the modelling framework typically also differ, creating an ensemble of models with complex connections. This parallel model development step is then  followed by a meta-analysis of the estimates of the key epidemic parameters at regular intervals, and the synthesized results are communicated to the public as well as used by policy makers \cite{department_of_health_and_social_care_r_number_methodology}.


Such a strategy has the benefits of robustifying inference on latent quantities, such as the much quoted effective reproduction number, here denoted $\Rt_t$, with regards to model misspecification, as well as countering  undue influence of artefacts connected to particular data sources. It also raises unresolved statistical issues on how best to formulate criteria for including suitable models in the ensemble, and how to weight a set of inter-connected models in the final estimate. Currently the UK government website lists seven academic groups and eight models which contribute to an ensemble \cite{department_of_health_and_social_care_r_number_methodology}, and pooled estimates are produced via a random effects meta-analysis in which all models are given equal weight \cite{maishman_statistical_2021}.  From a statistical perspective, using weights based on some measure of short-term predictive performance for each model would be a natural alternative, but would be potentially challenging to operationalize, as it requires all models to produce comparable predictive outputs;  see \cite{bracher_evaluating_2021} for a discussion of forecast evaluation metrics. Hence, meta-analysing results from an ensemble of models, while attractively operationally simple, results in overall estimates with unclear statistical properties, due to  uncertainty not being fully propagated. 

\subsection{Building a full joint model}

At the other end of the spectrum, one could strive to develop a single ``uber-model'' which contains all latent quantities of interest and a comprehensive set of data sources. The corresponding full joint posterior distribution of all the parameters of interest could then be derived simultaneously, using the paradigm of Bayesian graphical models. While theoretically optimal (provided the model is well specified) a full joint model is particularly challenging in a fast moving epidemic situation because it requires expansion, adaptation and revision each time a new piece of information needs to be integrated, whether it is new data source (e.g. presence of SARS-CoV-2 in geo-localised wastewater data) or a new policy or intervention influencing the structure of the model (e.g. vaccination).

Such increasing model complexity is accompanied by a heightened risk of misspecification and conflict between sources of information. This makes it hard to understand how different data sources are balanced in their contribution to the overall results, and difficult to track the influence of the assumptions made on how information is shared \cite{presanis_conflict_2013}, for example across space and time. Moreover, inevitable errors and quirks that occur in real time epidemiological data can result in contamination of inference whereby misspecification in one part of the model adversely affects analysis in another part. This may be hard to diagnose and correct for. 
Computationally, fitting a full joint model  
requires more intricate and dedicated programming, though this can be mitigated if full modularity is embedded in the programming language; in Section~\ref{sec:comp_strategy} we will discuss such computational strategies with particular reference to the \texttt{Turing} language \citep{Ge2018-am}, one of the probabilistic programming languages we are using in the Lab. It is also likely to be computationally demanding as data accrues and the associated number of model parameters increases, requiring frequent fine tuning of inference algorithms. 

Nevertheless, joint modelling is indispensable for certain tasks. For example, the University of Cambridge Medical Research Council Biostatistics Unit (BSU) and Public Health England (PHE) model uses a deterministic age-structured compartmental model \cite{birrell_real-time_2021}, data on daily \gls*{covid} confirmed deaths, and published information on the risk of dying and the time from infection to death, as primary data sources to estimate the number of new \gls*{sarscov2} infections over time. 
Starting in the early months of 2020 from a pre-existing flu transmission model,  the BSU-PHE model has continually been adapted and complexified. The model now accounts for the ongoing immunisation programme, differential susceptibility to infection in each adult age group, and incorporates estimates of community prevalence from the Office of National Statistics \gls*{covid} Infection Survey \cite{pouwels2021community}.\footnote{\url{https://www.ons.gov.uk/surveys/informationforhouseholdsandindividuals/householdandindividualsurveys/covid19infectionsurvey}}

\subsection{Interoperability of models -- the middle ground}

Between building separate models and building a single full joint model, the Lab experience of the \gls*{covid} pandemic has motivated us to adopt a strategy of interoperability of models for disease surveillance. Our overarching goal is to ensure adaptability of models and modules so they can be repurposed as needed, while maintaining a consistent treatment of uncertainty, and ensuring our approach is as robust as possible. We see interoperability as a journey, and the case studies presented in Section \ref{sec:case studies}  are there to show the direction of travel, not the fully equipped arrival lounge.

A software engineering analogy is pertinent here:  we believe that there is benefit from moving from a ``parallel'' approach, in which several separate models are analysed simultaneously and then the results integrated by model averaging, towards a ``serial'' approach, where input and output components and loose chains of models are considered. 
In a straightforward serial process, one could use posterior output as input into the next model. However, doing this in a fully Bayesian manner can be as computationally demanding as a full uber-model. Furthermore, as we will illustrate in our first case study (Section~\ref{sec:debias}), there are instances where it is beneficial to cut feedback \cite{jacob_better_2017}. In other cases,  the serial process will be akin to approximate Bayesian melding where posterior outputs are approximated by a suitable parametric distribution \cite{goudie_joining_2019}.

\section{The many levels of interoperability}

Interoperability is driven by a desire to deliver timely and robust 
statistical inference to answer several related research questions on a common process. As such, interoperability affects many aspects of the statistical workflow.

\subsection{Model specification and inference}
\subsubsection{\textbf{Modelling.}}
\label{sec:modelling}
We consider there to be an important distinction between an underlying model for the scientific process of interest, $S$ say, with parameters $\theta$ whose interpretation does not depend on what data are available, and an
observation model for data $D$ given $S$, with parameters $\phi$.  Dawid \cite{dawid1985probability}
calls $\theta$ and $\phi$ {\it extrinsic} and {\it intrinsic} parameters, respectively. Making this distinction clarifies how a new data-source can be
added to an existing model without the need to re-build the model from scratch. In the current context, our inferential focus is on extrinsic 
parameters such as incidence, prevalence or the growth rate.  We note, however, that care is needed in defining precisely the inferential target to answer any particular question; for example, an unqualified reference to ``prevalence'' is open to multiple interpretations. In the case studies that follow, we use point prevalence defined as the number of individuals in the population who would be found to be PCR-positive if tested, averaged over a specified time interval (e.g.\ over a week for the debiasing model described in Section~\ref{sec:debiasing_model}).

Once latent quantities and scenarios of interest are specified, a common modelling framework is desirable for building each model. We have chosen to formulate our models within the flexible framework of Bayesian hierarchical models (BHM) as it brings to the foreground conditional independence assumptions between the quantities of interest (whether observables or not) and  encodes the probabilistic relationships between them. BHMs clarify information flows through the use of Directed Acyclic Graphs (DAG), key assumptions made on exchangeability and ways of borrowing information. They also enable inclusion of new data sources in the DAGs in a coherent manner.
    
Considerations of interoperable modularity suggest there are benefits in targeting the marginal distributions of core quantities that feature within multiple models. This is because inference from these marginal models can then be fed into multiple downstream analyses, providing a common and coherent representation of key parameters. 

\subsubsection{\textbf{Statistical inference.}}
Adopting a common inference framework for all models allows a unified interpretation of their results; here we choose to use Bayesian inference.  Additionally, it is opportune to adopt common ways to validate outputs including the type of metrics to use, e.g. the use of marginal predictive distributions on observables. Working within a Bayesian inference framework also enables the possibility of coherent propagation of uncertainty without resorting to the direct specification of a full joint model. This can be accomplished via Markov melding \cite{goudie_joining_2019}, which builds upon the ideas of Markov combination \cite{dawid_hyper_1993,massa_combining_2010} and Bayesian melding \cite{poole_inference_2000}. Markov melding can be used to join several Bayesian models that involve a common parameter, with the prior combined using a ``pooling function''.  Indeed, we use a form of Markov melding with product-of-expert pooling for the bias parameter in our core model, which combines two sources of data to infer the posterior distribution of debiased prevalence (see Section~\ref{sec:debias}). We also rely on a normal two-stage approximation in our last two case studies (Sections~\ref{sec:st_modelling} and \ref{sec:epimap_modelling}), which can be viewed as an approximation to Markov melding, as discussed in \cite{goudie_joining_2019}. As an alternative to Markov melding, we have also used cut posterior distributions \cite{jacob_better_2017}  to give more weight to an unbiased-by-design source of evidence within our interoperability framework, making sure that we validate the resulting posterior distribution against gold-standard data.   
\subsection{Computational strategy}
\label{sec:comp_strategy}
    
\subsubsection{\textbf{Modularity.}}
 It is often helpful to decompose a complicated model into smaller modules. Such an approach can be useful for both the full joint model approach and the interoperability approach. For the former, an elaborate joint model can be decomposed into smaller modules, e.g. different parts of the prior, the evolving dynamics of the latent epidemic, and the likelihood model for different data sources. In a Bayesian framework, this approach is sometimes called a ``recursive'' or ``two stage'' approach \cite{hooten_recursive_2019, lunn_twostage_2013}, and has been advocated as a way to enable computationally-efficient model exploration and cross-validation \cite{goudie_modelexplore_2015} and integrate different statistical software packages \cite{johnson_greater_2020}.

Such a modular approach has the benefit to make each modelling component easier to conceptually reason about, validate, and communicate. From an operational point of view, it is also easier to perform unit testing, debugging and identification of computational bottlenecks within small, self-contained modules. For example, one can perform the prior, posterior or conditional predictive checks for these components independently. In principle, such an approach could create a spectrum of choices between the full joint model approach and the interoperability approach.  With the modules, one can freely choose between building a joint model by assembling these components or treating them as separate models, perform inference, and connect them using a cut or melding mechanism discussed in previous sections. 
    
    \subsubsection{\textbf{Probabilistic programming.}}
 One powerful way of performing statistical inference in an automated and timely manner is probabilistic programming, which allows one to write models in a concise, modular, intuitive syntax and automate Bayesian inference by using generic inference strategies (e.g. Gibbs sampling, Hamiltonian Monte Carlo). This significantly speeds up iterating models\footnote{The process of specifying and estimating models, making it practicable to explore a range of models.} during a preliminary data analysis phase. However, most probabilistic programming languages lack native support for interoperability. In order to experiment with the modularisation principle, we quickly added an experimental  \texttt{module} feature 
 to the \texttt{Turing} language \citep{Ge2018-am}. Each specified \texttt{module} behaves like an independent model and allows us to perform all kinds of operations and diagnostics available to a full model. In addition, the programming language allows these \texttt{module}s to be combined into more complex models, similarly to assembling elemental probability distributions. Thus, this \texttt{module} feature allows us to break a highly complex model into many modular, reusable modules. We experimented with the \texttt{module} feature on the Epimap model of the local reproduction number \cite{teh2021}. The module mechanism enables us to implement interoperability between Epimap and the Debiasing model with a minimal amount of extra work (i.e. one author implemented this interoperability of Epimap and Debiasing models within a day). It also allowed us to spot and fix a computational bottleneck arising from distributional changes. We discuss more details of this example of interoperability in section \ref{sec:epimap_modelling}.

\subsection{Data deployment towards interoperable modelling} 
The underpinning data and its synchronisation is of paramount importance to ensure interoperable modelling, requiring meticulous data synthesis pipelines and high quality data curation and tracking. 
\subsubsection{\textbf{Ensuring consistency of data feeds.}}
To ensure that interoperable models provide a coherent and robust set of outputs, consistent, high quality data feeds must be applied. This is challenging when the generation of datasets are evolving rapidly, the data is frequently changing, and the downstream synthesis is continually being updated and modified. The complex datasets generated are typically being made available to a diverse user group resulting in data proliferation and redundancy. Parallel, superficially redundant data feeds may be maintained to increase resilience. Note that differing levels of information governance dependant on the proposed data use dictate the granularity of data available to specific users, requiring strict data management to ensure data synchronisation.  

Data curation processes may be performed at the source of the data and centrally for both 
basic data cleaning and sophisticated curation \cite{yu_veridical_2020}. The tracking of metadata describing the data granularity and data curation are important to assess outputs that are comparable. All of these factors mean that the same core data may be required to be provided to users in several different forms, with varied provision and annotation of metadata. The data selection requirements for our interoperable models may also vary, for example the level of granularity of age or geography may be different, and so the transparency of the data transformation is key to informing interoperability.

\subsubsection{\textbf{Reducing data redundancy.}}
To reduce data redundancy, ``Single-source-of-truth'' or ``Master Data Management principles'' could be applied to the provision of datasets. These principles suggest that only a single, master copy of each dataset is maintained, and that datasets are combined into for example a data warehouse by linking rather than duplication. The deployment of robust ETL (Extract-Transform-Load) processes that implement the differing business logic (e.g. data transformations) to capture and integrate data from multiple feeds into a single data repository facilitates downstream consistent data feeds, data synchronisation and reduces redundancy.  In this manner, the ETL facilitates the definition of the data transformation and processes required to manage rapidly evolving data and its underlying structure.  The data warehouse is then a stable research ready dataset that facilites data releases and snapshots of the primary data, e.g. \gls*{covid} test results, to specified users.

\subsubsection{\textbf{Improving reproducibility.}}
Reproducibility is the principle that all policy decisions or publication should be based on a data analysis output that can be replicated. Reproducible, reliable, and transparent results comprise one of the key ingredients of the data life science framework for veridical data science proposed in \cite{yu_veridical_2020}.  To achieve reproducibility, both input data and software code must be versioned, and each version must be retrievable at a later date. Code versioning is easily achieved using a source code management tool such as Git \cite{git}. Data versioning can be achieved through a variety of methods, depending on the stability of the data structures. For static data structures, temporal tables (or manually timestamped records) provide the ability to query data as of a specific time. The normalised data structures described above promote static data structures; when new fields are required, a new table is used instead of adding fields to existing tables. When changes in schemas are unavoidable, regular data dumps with named versions may be more appropriate.

\section{Case studies}
\label{sec:case studies}

The case studies we present are chosen to illustrate some of the benefits and issues relating to interoperability that the Lab has encountered in its work. The concept of interoperability first crystallised through our work on debiased prevalence \cite{nicholson_local_2021}. In Section~\ref{sec:adjust_ascertainment_bias} we describe this work and demonstrate how careful information synthesis (in this case cutting feedback) between data modules helped us to robustify prevalence estimation.
In Section~\ref{sec:SIR_debiased} we demonstrate how we can coherently integrate and propagate the uncertainty of resulting estimates of debiased prevalence into a compartmental epidemic model.

In the final two case studies in Sections~\ref{sec:st_modelling} and \ref{sec:epimap_modelling}, we provide additional examples of model interoperability, showing different methods of inputting debiased prevalence into other models to answer new questions. In each case study, we outline some additional advantages (e.g. computational ease) but also new questions that arise in the process (e.g. appropriate handling of uncertainty in the prevalence estimates or how to link models built for different time scales).

Overall, the case studies demonstrate how analyses in a demanding, fast-moving, public health context can be operationalised within a framework that flexibly combines a set of independent analysis modules. 


\subsection{The data}
We first describe the data used across the case studies.  
\subsubsection{\textbf{Randomized surveillance data.}} These record $\nr$ positive tests out of $\Nr$ total subjects tested. \label{sec:rand_data_intro} The \gls*{react} study is a nationally representative prevalence survey of \gls*{sarscov2} based on repeated \gls*{pcr} tests of cross-sectional samples from a representative subpopulation defined through stratified random sampling from England's National Health Service patient register \citep{riley2020community}.

\subsubsection{\textbf{Targeted surveillance data.}} These record $\nt$ positive tests of $\Nt$ total subjects tested. Pillar 1 tests comprise \textit{``all swab tests performed in \gls*{phe} labs and \gls*{nhs} hospitals for those with a clinical need, and health and care workers''}, and Pillar 2 is defined as \textit{``swab testing for the wider population''} \cite{department_of_health_and_social_care_uk_covid-19_nodate}. Pillar 1+2 testing has more capacity than \gls*{react}, but the protocol incurs ascertainment bias as those at higher risk of being infected are more likely to be tested, such as front-line workers, contacts traced to a \gls*{covid} case, or the sub population presenting with \gls*{covid} symptoms, such as loss of taste and smell \cite{department_of_health_and_social_care_uk_covid-19_nodate}. The ascertainment bias potentially varies over the course of the pandemic as the testing strategy and capacity changes. We exclude lateral flow tests and use exclusively test data from Pillar 1+2 \gls*{pcr} tests. 

\subsubsection{\textbf{Population metadata.}} We enrich the testing data by population characteristics related to the following measures of ethnic diversity and socio-economic deprivation in each local area: 
\begin{itemize}
    \item \emph{Ethnic diversity.} The proportion of BAME (Black, Asian and Minority Ethnic) population is retrieved from the 2011 Census. 
    \item \emph{Socio-economic deprivation.} The 2019 Index of Multiple Deprivation (IMD) score is retrieved from the Department of Communities and Local Governments \cite{england_indices_of_deprivation_2019}.
    IMD is a composite index calculated at Lower Super Output Area level (LSOA) and based on several domains representing deprivation in income, employment, education, crime, housing, health and environment. For other geographies, e.g. Lower Tier Local Authority (LTLA) which we use in the case studies, the IMD is obtained as the population weighted average of the corresponding LSOAs, using the 2019 mid-year population counts.
\end{itemize}

\subsubsection{\textbf{Commuter flow data.}} \label{sec:commuter_flow} As one of its data inputs, the Epimap model \cite{teh2021} uses commuting flow data from the 2011 Census.\footnote{https://census.ukdataservice.ac.uk/use-data/guides/flow-data.aspx} After preprocessing, these data are used to create a flux matrix $F$ that determines how transmission events occur within and between \glspl*{ltla}; see \cite{teh2021} for full details. In Section~\ref{sec:interpretation_via_data_sync} we will discuss the relationship between flux matrix $F$ and estimates of $\Rt_\subiw$.

\subsection{Estimating and adjusting for ascertainment bias in $\pill$ data}
\label{sec:adjust_ascertainment_bias}

We now summarise our debiasing model, which combines targeted surveillance data with randomized surveillance data to obtain local estimates of prevalence. Full details can be found in \cite{nicholson_local_2021} along with accompanying R code.\footnote{\url{https://github.com/alan-turing-institute/jbc-turing-rss-testdebiasing}} 
\label{sec:debias}
\subsubsection{\textbf{Debiasing model.}}
\label{sec:debiasing_model}

The \gls*{react} data provide accurate but relatively imprecise estimates of prevalence at the \gls*{phe} region level (i.e.\ coarse scale). Note that \gls*{react} total test counts $U$ tend to be much smaller than $\pill$ test counts $N$, with $U/N$ of order $10^{-2}$. The \gls*{react} data likelihood for the PCR-positive prevalence proportion $\prevprop$ is
\begin{eqnarray}
\label{eq:rand_lik_regtime_dep}
	\mathbb{P}(\nrofN \mid \prevprop) &=& \mathrm{HyperGeometric}(\nr \mid \popsize, \prevprop\popsize, \Nr)\ ,\ \ \ \text{for }\prevprop\popsize\in \mathbb{Z}
\end{eqnarray}
based on observing $\nr$ positive tests out of a total of $\Nr$ randomly allocated in a population of known size $\popsize$; our inference under \eqref{eq:rand_lik_regtime_dep} and \eqref{eq:pillar_2_lik_mod_v1} is based on a latent integer number of infected $\prevprop\popsize$ (see \cite{nicholson_local_2021} for details). Note that, for simplicity, we are ignoring the stratified sampling design in likelihood \eqref{eq:rand_lik_regtime_dep}.

In contrast, test positivity rates in Pillar 1+2 data are strongly biased upwards relative to the population prevalence proportion, as the testing is directed at the higher risk population (e.g. symptomatics, frontline workers). We show how careful modelling of the ascertainment process allows us to estimate prevalence accurately, and with good precision, even at a fine-scale level such as LTLA.

We introduce the following causal model for the observation of $\ntofN$ positive targeted (e.g.\ Pillar 1+2) tests:
\begin{eqnarray}
    \nonumber
    \mathbb{P}(\ntofN \mid \prevprop, \delta, \nu) &=& \mathrm{Binomial}\left(\nt \mid \prevprop \popsize,\ \Pp(\Tes \mid \Infd)\right)\ \\   
    \label{eq:pillar_2_lik_mod_v1}
    &&\times\ \mathrm{Binomial}(\Nt - \nt \mid (1 - \prevprop)\popsize,\ \Pp(\Tes \mid \text{Not}\  \Infd))
\end{eqnarray}
where $\delta$ and $\nu$ parameterise (on log odds scale) the binomial success probabilities $\Pp(\Tes \mid \Infd)$ and $\Pp(\Tes \mid \text{Not}\  \Infd)$: 
\begin{eqnarray}
    \delta &:=& \log\left( \frac{\mathrm{Odds}(\Tes \mid \Infd)}{\mathrm{Odds}(\Tes \mid \text{Not}\  \Infd)}\right)\\
    \nu &:=& \log \mathrm{Odds} (\Tes \mid \text{Not}\  \Infd)\ .
\end{eqnarray}
By causal, we mean that we explicitly take into account the way the Pillar 1+2 data was generated by inferring, and conditioning  in \eqref{eq:pillar_2_lik_mod_v1}, on the probability of individuals in the population being tested. We provide a detailed description of the model in \cite{nicholson_local_2021}. The parameter requiring careful treatment is $\delta$, i.e.\ the log odds ratio of being tested in the infected versus the non-infected subpopulations. The other parameter, $\nu$, is directly estimable from the targeted data, with $\hat{\nu}:= \text{logit}[(N-n)/\popsize]$ acting as a precise estimator with little bias when prevalence is low. We use $\hat{\nu}$ as a plug-in estimator in likelihood \eqref{eq:pillar_2_lik_mod_v1}; this is for computational convenience -- if we knew $\pi$ we would instead estimate $\hat{\nu}:= \text{logit}[(N-n)/\{\popsize(1-\pi)\}]$. While allowing efficient computation, heuristically estimating $\nu$ can contribute to model misspecification, a point to which  we return in Section~\ref{sec:modularized}.
 
\begin{figure}[ht]
	\vspace{.4cm}
\begin{center}
    \begin{tikzpicture}
		\node[obs] (pillar12) {$\pilln$};
		\node[obs, above=of pillar12] (react) {$\reactn$};
		\node[latent, left=of react] (prevprop) {$\prevprop$};
		\node[latent, left=of pillar12] (p12bias) {$\biasparpill$};
		\node[const, left=of p12bias] (p12biasprior) {$p_{\text{flat}}(\biasparpill)$};
		\node[const, left=of prevprop] (prevpropprior) {$p(\prevprop)$};
		\node[const, above=.75cm of prevprop, text width=6cm] (coarse) {(a) Coarse scale (e.g.\ PHE region)\\ 
		\hspace{.45cm} Learn $\biasparpill$ via ``cut'' model};
		\edge {prevprop} {react}; 
		\edge[dashed] {prevprop} {pillar12}; 
		\edge {p12biasprior} {p12bias}; 
		\edge {p12bias} {pillar12}; 
		\edge {prevpropprior} {prevprop}; 
	\end{tikzpicture}
	\hspace{2cm}
    \begin{tikzpicture}
		\node[obs] (pillar12) {$\pilln$};
		\node[latent, left=of react] (prevprop) {$\prevprop$};
		\node[const, above=.75cm of prevprop, text width=6cm] (fine) {(b) Fine scale (e.g.\ LTLA)\\ 
		\hspace{.45cm} Learn $\prevprop$ from $\pill$};
		\node[latent, left=of pillar12] (p12bias) {$\biasparpill$};
		\node[const, left=of prevprop] (prevpropprior) {$p(\prevprop)$};
		\node[const, left=of p12bias] (p12biasprior) {$p_{\text{DD}}(\biasparpill)$};
		\edge {prevprop} {pillar12}; 
		\edge {p12bias} {pillar12}; 
		\edge {p12biasprior} {p12bias}; 
		\edge {prevpropprior} {prevprop}; 
	\end{tikzpicture}
\end{center}
\vspace{-.25cm}
	\caption{Models for debiasing. (a) Cut model where the dashed line represents cutting feedback from $\pill$ to $\prevprop$. (b) Data-dependent prior  $p_{\text{DD}}(\biasparpill)$ has been created from cut posterior for $\biasparpill$ from stage (a), and $\pill$ used to infer $\prevprop$ at the local (\gls*{ltla}) level.}
	\label{fig:cut_model}
\end{figure}

\subsubsection{\textbf{Constraining information flow for robust estimation of ascertainment bias $\delta$.}}
\label{sec:modularized}
In Figure~\ref{fig:cut_model}(a), the joint posterior distribution (without cutting any information flow) is
\begin{align}
    \label{eq:uncut_model}
    p(\prevprop, \biasparpill \mid \reactn, \pilln) &= p(\prevprop \mid \reactn, \pilln) \times p(\biasparpill \mid \prevprop, \pilln)\ .
\end{align}
This provides optimal, coherent inference when the model space contains the true data generating mechanism. However, if the  $\pill$ likelihood \eqref{eq:pillar_2_lik_mod_v1} is misspecified, then inference on $\prevprop$, and hence on $\biasparpill$, can be adversely affected. In the current context, the consequences of misspecification are particularly severe because, conditional on $\biasparpill$, the relative sample sizes lead to the $\pill$ likelihood \eqref{eq:pillar_2_lik_mod_v1} typically containing far more information on $\prevprop$ than the $\gls*{react}$ likelihood \eqref{eq:rand_lik_regtime_dep}.

\begin{figure}
	\begin{adjustbox}{max totalsize={\textwidth}{\textheight},center}
	    \includegraphics{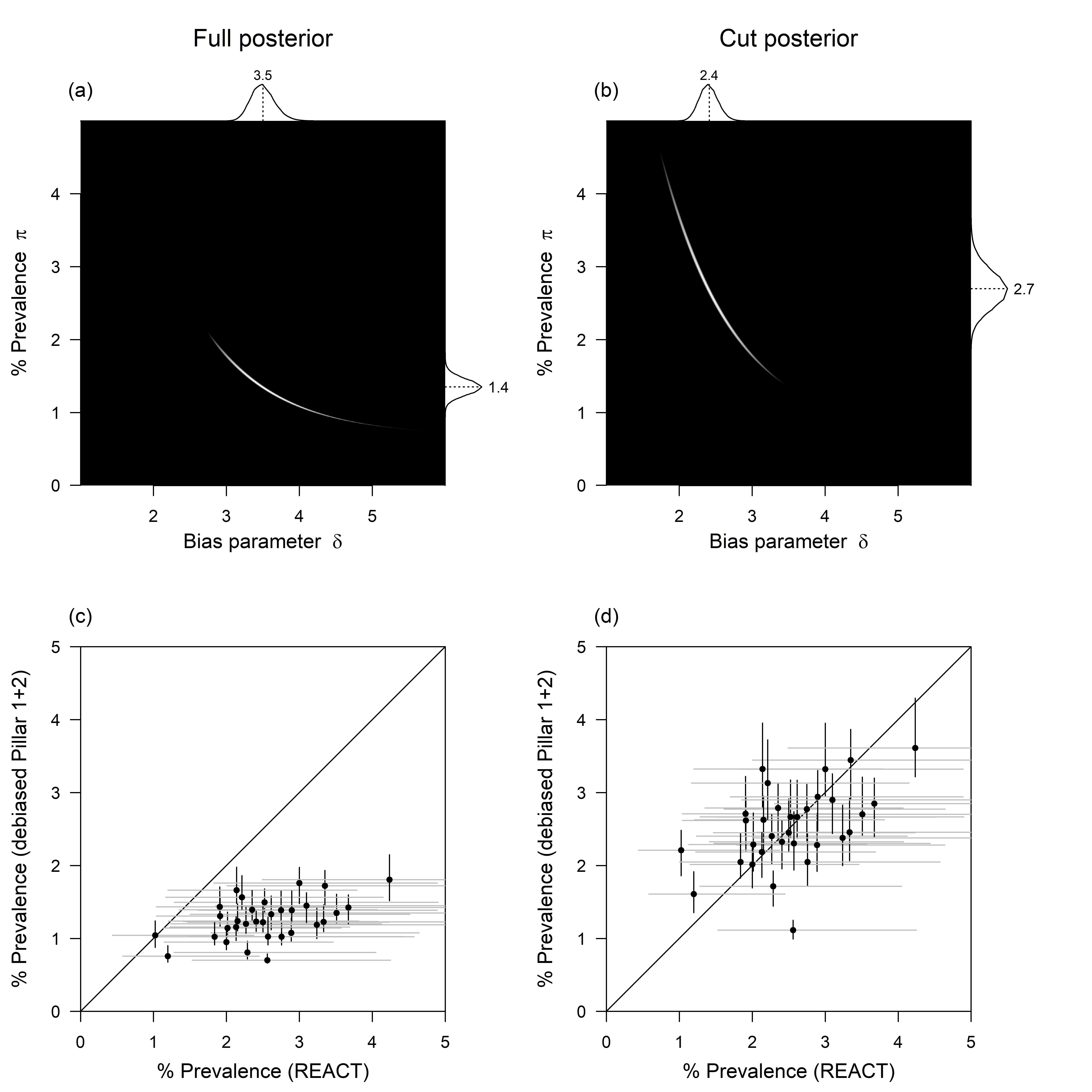}
	\end{adjustbox}
	\caption{Comparison of full vs cut posterior inference for bias parameter $\biasparpill$. (a) \textbf{Full} joint log posterior --- logarithm of equation \eqref{eq:uncut_model}. (b) \textbf{Cut} joint log posterior --- logarithm of equation \eqref{eq:cut_model}. (c) \gls*{ltla}-level debiased \% prevalence based on \textbf{full} posterior (vertical) vs gold-standard \gls*{react} \% prevalence (horizontal), one point per \gls*{ltla}. (d) \gls*{ltla}-level debiased \% prevalence based on \textbf{cut} posterior (vertical) vs gold-standard \gls*{react} (horizontal). In panels (a-b) the marginal posterior distributions are plotted at the panel top and right edges, with marginal means labelled. 
    Inference is based on \gls*{react} and $\pill$ data for London and its constituent \glspl*{ltla} for \gls*{wc} 14th Jan 2021. The gold-standard \gls*{ltla}-level estimates used for validation on the horizontal axes in (c-d) are based on \gls*{react} round 8 aggregated data (6th-22nd Jan 2021).}\label{fig:cut_vs_full}
\end{figure}

With this in mind, we use a cut posterior distribution, as described in \cite{jacob_better_2017}:     
\begin{align}
    \label{eq:cut_model}
    p^{\text{cut}}(\prevprop, \biasparpill \mid \reactn, \pilln) &:= p(\prevprop \mid \reactn)
    \times p(\biasparpill \mid \prevprop, \pilln)
\end{align}
where the first distribution on the RHS of  \eqref{eq:cut_model} is no longer conditioning on the $\pilln$ from $\pill$. Switching from model \eqref{eq:uncut_model} to \eqref{eq:cut_model} ``cuts feedback'' from $\pill$ to $\prevprop$ in inference on $(\prevprop, \biasparpill)$. Figure~\ref{fig:cut_vs_full}(a-b) compares the joint full posterior \eqref{eq:uncut_model} with the joint cut posterior \eqref{eq:cut_model} for $\pill$ and \gls*{react} data  gathered in London during \gls*{wc} 14th Jan 2021. In this week the $\pill$ data were $\nt\text{ of }\Nt=$ 60,749 of 326,986 and the \gls*{react} data were $\nr\text{ of }\Nr=$ 101 of 3,778, i.e.\ the $\pill$ data have 87 times as many tests as \gls*{react} that week. 

The joint posteriors in Figure~\ref{fig:cut_vs_full}(a-b) have clearly quite different support. The mean (95\% CI) for $\biasparpill$ under the full posterior is \mbox{3.5
\unskip}\ (\mbox{3.2
\unskip}-\mbox{3.9
\unskip}) whilst under the cut posterior it is  \mbox{2.4
\unskip}\ (\mbox{2.2
\unskip}-\mbox{2.7
\unskip}). Which model, full or cut, is estimating $\biasparpill$ more accurately?  
First, note that in the cut posterior \eqref{eq:cut_model} the marginal distribution of the prevalence proportion $\prevprop$ depends only on the \gls*{react} data, 
\begin{align}
    \label{eq:cut_model_pi_marg}
    p^{\text{cut}}(\prevprop \mid \reactn, \pilln) &= p(\prevprop \mid \reactn)\ ,
\end{align}
and that the maximum likelihood estimator for $\prevprop$ based on model $\mathbb{P}(\nrofN \mid \prevprop)$ at \eqref{eq:rand_lik_regtime_dep} is approximately unbiased, since \gls*{react} is a designed, randomised study. We say that the estimator is approximately unbiased as, for simplicity, we do not account for the stratified sampling design nor for the  non-response ($>75\%$), which we assume is non-informative. Thus, under a weakly informative prior $p_{\text{flat}}(\prevprop)$, the cut-posterior $\prevprop$-marginal mean (95\% CI) of \mbox{2.7
\unskip}\ (\mbox{2.2
\unskip}-\mbox{3.2
\unskip}) estimates $\prevprop$ reasonably accurately. However, the full-posterior $\prevprop$-marginal mean of \mbox{1.4
\unskip}\ (\mbox{1.1
\unskip}-\mbox{1.6
\unskip}) is quite different, suggesting that the $\pill$ causal testing model at \eqref{eq:pillar_2_lik_mod_v1} is misspecified and, having a relatively large amount of data, swamps the accurate information in the \gls*{react} data. 
We can attribute part of the misspecification in this case to the bias in the estimator $\hat{\nu}= \text{logit}[(N-n)/\popsize]$ introduced in Section~\ref{sec:debiasing_model}. The bias is larger for high prevalence and has a greater effect when the $\pill$ sample size is large (both of which apply to the illustrative week examined of London w/c 14th Jan 2021, where prevalence is estimated at 2.7\% and $\Nt=326,986$).  We further compare and contrast inference under the full and cut posteriors after having applied them to estimate local prevalence in Section~\ref{sec:cross_sectional_local_prevalence}.

As a further point of note on the use of cut posterior distributions to robustify inference, many epidemiological models make use of knowledge of generation intervals, serial intervals, and incubation periods. These are typically estimated from small-scale but direct studies of transmissions, producing broad confidence/credible intervals reflecting the inherent uncertainties (e.g.\  \citep{bi2020epidemiology}). If these are used as priors for generation/serial intervals or incubation periods for a large scale epidemiological model such as Epimap (\cite{teh2021}; see Section \ref{sec:epimap_modelling}), the lower quality yet larger amounts of information in the large scale data (e.g.\ test counts in case of Epimap, but can also include hospitalisation and death counts) can easily overwhelm the priors. Instead a common approach (besides Epimap, see also \cite{brauner_inferring_2021}) is to draw multiple samples from the priors of these quantities, compute the posterior predictive distribution for each sample, then aggregate across prior samples to capture the uncertainties over the generation/serial interval and incubation period. This can be equivalently viewed as nested Monte Carlo estimation for a cut posterior distribution. 


\subsubsection{\textbf{Inferring cross-sectional local prevalence.}} \label{sec:cross_sectional_local_prevalence}

We specify a data-dependent prior $p_{\text{\gls*{dd}}}(\biasparpill_\subjw)$ for each week $\weeksub$ in each region $\phesub$, based on the posterior distribution at \eqref{eq:cut_model}. We begin by approximating the marginal posterior distribution for $\biasparpill_\subjw$ with a moment-matched Gaussian based on the cut posterior:
\begin{align}
\label{eq:eb_prior_first}
    p_{\text{\gls*{dd}}}(\biasparpill_\subjw) &:= \mathrm{Normal}(\delta_{\phesub,w}\mid \hat{\mu}_\subjw, \hat{\tau}^2_\subjw)\\
    &\approx p^{\text{cut}}(\biasparpill_\subjw \mid \reactnjw, \pillnjw).
\end{align}
For example, in the case of the marginal density at the top of Figure~\ref{fig:cut_vs_full}(b), we specify $p_{\text{\gls*{dd}}}(\biasparpill_\subjw)$ by setting $\hat{\mu}_\subjw=\mbox{2.4
\unskip}$ and $\hat{\tau}_\subjw=\mbox{0.1
\unskip}$ for London \gls*{wc} 14th Jan 2021. For comparison's sake, we will also consider later an alternative data-dependent prior  $p_{\text{\gls*{dd}}}(\biasparpill_\subjw)$ which uses the full posterior based on \eqref{eq:uncut_model}, leading to different downstream inference.
 
 While this approach provides a prior for weeks at which both $\pill$ and \gls*{react} data are available (since we are able to estimate $\biasparpill$), we also wish to interpolate and/or extrapolate information on $\biasparpill$ to weeks at which \gls*{react} data are unavailable (e.g.\ between sampling rounds). We achieve this by introducing a smoothing component into a product-of-experts prior (\cite{hinton_training_2002}; details in Appendix~\ref{sec:smoothing_prior_for_delta}), thereby allowing us to specify independent priors $p_{\text{\gls*{dd}}}(\biasparpill_\subjw)$ of the form \eqref{eq:eb_prior_first} for all weeks, including those without \gls*{react} data. 

Having specified a prior at the coarse-scale regions $\phesub$, we proceed to perform full Bayesian inference at a fine-scale \gls*{ltla} $\ltlasub$ using the prior from its corresponding region, $p_{\text{\gls*{dd}}}(\delta_{\phesub[\ltlasub],\weeksub})$. 
We plot cross-sectional \% prevalence posterior medians (with 95\% posterior CIs) on the vertical axes of Figure~\ref{fig:cut_vs_full}(c) and (d) for the full and cut models respectively, with each point corresponding to the estimated \% prevalence for one LTLA in London for \gls*{wc} 14th Jan 2021. Observe in Figure~\ref{fig:cut_vs_full}(c-d) that the horizontal \gls*{react} CIs are relatively broad, compared to the vertical debiased $\pill$ CIs, exemplifying the relatively large amount of information in the $\pill$ data if $\biasparpill$ can be inferred. 

The results in Figure~\ref{fig:cut_vs_full}(c-d) are also highly informative for our discussion of inference under the full vs cut models, because we are able to validate against ``gold-standard'' \gls*{ltla}-level randomised surveillance data aggregated across round 8 of the \gls*{react} study (6th-22nd Jan 2021). On the horizontal axis, we plot the \gls*{react} unbiased prevalence estimates (with 95\% exact binomial CIs). Compared to the accurate \gls*{react} estimates, the debiased prevalence estimates based upon the full posterior are biased downwards -- the estimated bias across the \glspl*{ltla} plotted is \mbox{-1.28
\unskip}\% (SE = \mbox{0.11
\unskip}\%). In contrast, the debiased prevalence estimates based on the cut posterior appear to be accurate, having estimated bias of \mbox{-0.01
\unskip}\% (SE = \mbox{0.11
\unskip}\%). 

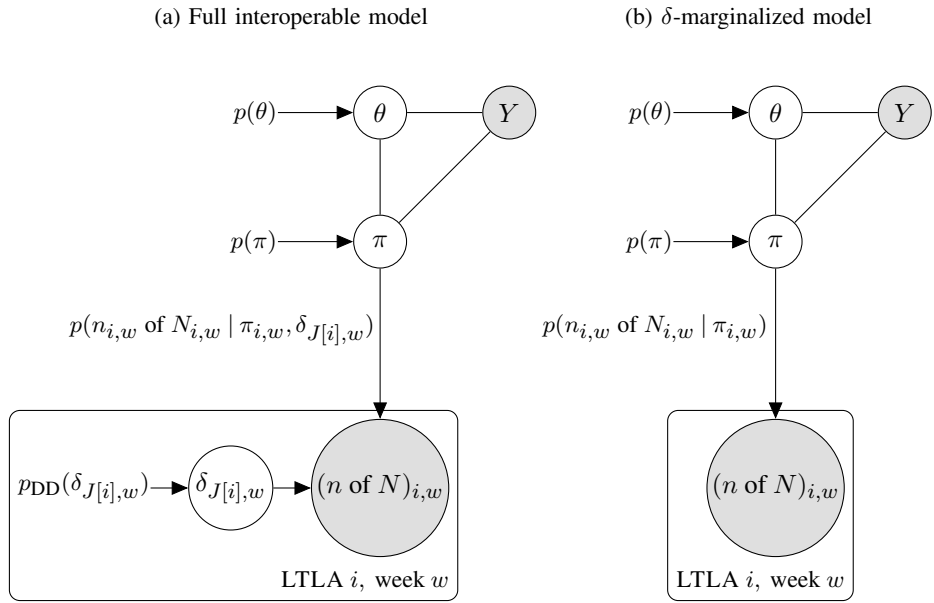
\begin{figure}[h]
	\vspace{.4cm}
\begin{center}
    \begin{tikzpicture}
		\node[obs] (pillar12) {$(\pilln)_\subiw$};
		\node[latent, left=of pillar12, xshift=.5cm] (p12bias) {$\biasparpill_{\phesub[\ltlasub],\weeksub}$};
		\node[latent, above=of pillar12, yshift=1cm] (prevprop) {$\prevprop$};
		\node[latent, above=of prevprop] (theta) {$\theta$};
		\node[obs, right=of theta] (y) {$Y$};
		\node[const, above=.75cm of theta, text width=6cm] (fine) {(a) Full interoperable model};
		\node[const, left=of prevprop] (prevpropprior) {$p(\prevprop)$};
		\node[const, left=of p12bias, xshift=.5cm] (p12biasprior) {$p_{\text{DD}}(\biasparpill_{\phesub[\ltlasub],\weeksub})$};
		\node[const, left=of theta] (thetaprior) {$p(\theta)$};
				\node[const, below=of prevprop, xshift=-2.1cm, yshift=.4cm] (distlab) {$p(\text{$\nt_\subiw$\ of\ $\Nt_\subiw$} \mid \prevprop_\subiw, \biasparpill_{\phesub[\ltlasub],\weeksub})$};

		\edge {prevprop} {pillar12}; 
		\edge[-] {theta} {y}; 
		\edge[-] {prevprop} {y}; 
		\edge[-] {theta} {prevprop}; 
		\edge {thetaprior} {theta}; 
		\edge {p12bias} {pillar12}; 
		\edge {p12biasprior} {p12bias}; 
		\edge {prevpropprior} {prevprop}; 
		\plate {import} {(p12biasprior)(pillar12)(p12bias)} {{$\text{\gls*{ltla}\ }i,\ \text{week\ }w$}};
	\end{tikzpicture}
	\hspace{-1cm}
    \begin{tikzpicture}
		\node[obs] (pillar12) {$(\pilln)_\subiw$};
		\node[latent, above=of pillar12, yshift=1cm] (prevprop) {$\prevprop$};
		\node[latent, above=of prevprop] (theta) {$\theta$};
		\node[obs, right=of theta] (y) {$Y$};
		\node[const, below=of prevprop, xshift=-1.6cm, yshift=.4cm] (distlab) {$p(\text{$\nt_\subiw$\ of\ $\Nt_\subiw$} \mid \prevprop_\subiw)$};
        \node[const, above=.75cm of theta, text width=6cm, xshift=1cm] (fine) {(b) $\biasparpill$-marginalized model};
		\node[const, left=of prevprop] (prevpropprior) {$p(\prevprop)$};
		\node[const, left=of theta] (thetaprior) {$p(\theta)$};
		\edge {prevprop} {pillar12}; 
		\edge[-] {theta} {y}; 
		\edge[-] {prevprop} {y}; 
		\edge[-] {theta} {prevprop}; 
		\edge {thetaprior} {theta}; 
		\edge {prevpropprior} {prevprop}; 
		\plate {import} {(pillar12)} {{$\text{\gls*{ltla}\ }i,\ \text{week\ }w$}};
	\end{tikzpicture}
\end{center}
\vspace{-.25cm}
	\caption{Interoperable interface of debiasing output with a model parameterised by $\theta$ and with other data $Y$. (a) Full interoperable model. (b) Collapsed representation, in which the full model has been marginalized wrt $\biasparpill$. We use $\prevprop$ to denote the entire spatiotemporal collection of prevalence proportions $\prevprop_{1:I,1:W}$.}
	\label{fig:marginal_approx_lik_graph}
\end{figure}

\subsubsection{\textbf{General interoperability between models.}} 
The graph in Figure~\ref{fig:marginal_approx_lik_graph}(a)  relates the debiasing model (in the $\ltlasub,\weeksub$ plate) to another arbitrary model with parameters $(\prevprop, \theta)$ and data/covariates $Y$ (with $Y$ not containing $\pill$). One approach to fitting Figure~\ref{fig:marginal_approx_lik_graph}(a)'s model would be to perform full Bayesian inference directly, i.e.\ sampling from $(\theta, \prevprop, \biasparpill)$. However, as will be illustrated with the SIR model in Section~\ref{sec:SIR_debiased}, we can reduce the computational complexity by first marginalising with respect to $\delta_{\phesub[\ltlasub],w}$ yielding the marginal likelihood:
\begin{align}
\label{eq:marg_likelihoods}
    p(\text{$\nt_\subiw$\ of\ $\Nt_\subiw$} \mid \prevprop_\subiw)
    &= \int  p(\text{$\nt_\subiw$\ of\ $\Nt_\subiw$} \mid \prevprop_\subiw ,  \delta_{\phesub[\ltlasub],\weeksub})\ p_{\text{\gls*{dd}}}(\delta_{\phesub[\ltlasub],\weeksub})\ d\delta_{\phesub[\ltlasub],\weeksub}\ 
\end{align}
which is an unnormalized function of $\prevprop_{\phesub,w}$ encapsulating all information on $\prevprop_{\phesub,w}$ that results from observing $\nt_\subiw$\ of\ $\Nt_\subiw$ positive $\pill$ tests. As illustrated in Figure~\ref{fig:marginal_approx_lik_graph}(b), we then need sample only $(\theta, \prevprop)$ in the $\biasparpill$-marginalised interoperable model.


While the marginal likelihood $p(\text{$\nt_\subiw$\ of\ $\Nt_\subiw$} \mid \prevprop_\subiw)$ at \eqref{eq:marg_likelihoods} can easily be evaluated point-wise, it does not have a closed parametric form. We can further simplify inference at the interface between models by approximating (up to a multiplicative constant) the marginal likelihood with a parametric distribution. A Gaussian moment-matched approximation on log odds scale (Figure~\ref{fig:marginal_approx_lik_graph}(b)), 
\begin{align}
    \label{eq:gaussmomlik}
    \hat{p}(\text{$\nt_\subiw$\ of\ $\Nt_\subiw$} \mid  \prevprop_\subiw) &\overset{\prevprop_\subiw}{\propto} \text{Normal}(\text{logit}\ \hat{\prevprop}_\subiw \mid \text{logit}\ \prevprop_\subiw, \hat{\sigma}^2_\subiw)\ ,
\end{align}
is a natural choice here because it provides an empirically good fit and also integrates conveniently with methods and software for hierarchical generalised linear models with logit link function, as we shall see in Section~\ref{sec:st_modelling}. This approach of summarising a module with a moment-matched Gaussian distribution for use as an approximate likelihood term in subsequent modules has been previously widely used in simpler contexts, including evidence syntheses \cite{welton_evsyn_2012} and more general hierarchical models \cite{daniels_twostage_1998, lunn_twostage_2013}.

Depending on the context, either one of \eqref{eq:marg_likelihoods} or \eqref{eq:gaussmomlik} may be preferred. Using the exact marginal likelihood at \eqref{eq:marg_likelihoods} avoids making a Gaussian approximation, but \eqref{eq:marg_likelihoods} can be computationally unwieldy as it is a mass function on integer prevalence $\prevprop \popsize$. The Gaussian approximated marginal likelihood at \eqref{eq:gaussmomlik} is often more  computationally convenient to integrate with other models. We use \eqref{eq:marg_likelihoods} in Section~\ref{sec:SIR_debiased} and we use \eqref{eq:gaussmomlik} in Sections~\ref{sec:st_modelling} and \ref{sec:epimap_modelling}.


\subsection{\textbf{Interoperability with an SIR model.}} \label{sec:SIR_debiased} 
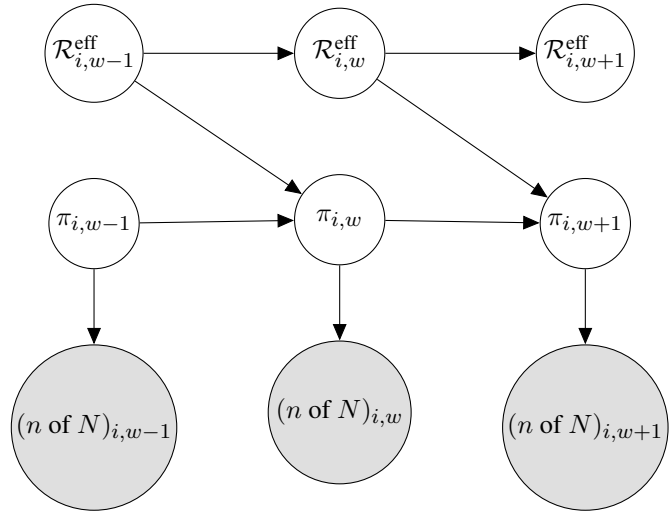
\begin{figure}
	\vspace{.4cm}
\begin{center}
		\begin{tikzpicture}
		\def\xlab{1cm}
		\def\ylab{0cm}
		\def\zlab{.7cm}
		\def\minsize{1.2cm}
		\def\minsizen{1.9cm}
		\node[latent,minimum size=\minsize] (rwminus1) {$\Rt_{\ltlasub,\weeksub - 1}$};
		\node[latent,minimum size=\minsize, right= of rwminus1, xshift=\xlab] (rw) {$\Rt_{\ltlasub,\weeksub}$};
		\node[latent,minimum size=\minsize, right= of rw, xshift=\xlab] (rwplus1) {$\Rt_{\ltlasub,\weeksub + 1}$};
		\node[latent,minimum size=\minsize, below= of rwminus1, yshift=\ylab] (piwminus1) {$\prevprop_{\ltlasub,\weeksub - 1}$};
		\node[latent,minimum size=\minsize, below= of rw, yshift=\ylab] (piw) {$\prevprop_{\ltlasub,\weeksub}$};
		\node[latent,minimum size=\minsize, below= of rwplus1, yshift=\ylab] (piwplus1) {$\prevprop_{\ltlasub,\weeksub+1}$};
		\node[obs,minimum size=\minsizen, below= of piwminus1] (nwminus1) {$(\pilln)_{\ltlasub,\weeksub-1}$};
		\node[obs,minimum size=\minsizen, below= of piw] (nw) {$(\pilln)_{\ltlasub,\weeksub}$};
		\node[obs,minimum size=\minsizen, below= of piwplus1] (nwplus1) {$(\pilln)_{\ltlasub,\weeksub+1}$};
		\edge {rwminus1} {rw};
		\edge {rw} {rwplus1};
		\edge {piwminus1} {piw};
		\edge {piw} {piwplus1};
		\edge {rwminus1} {piw};
		\edge {rw} {piwplus1};
		\edge {piwminus1} {nwminus1};
		\edge {piwplus1} {nwplus1};
		\edge {piw} {nw};
		\end{tikzpicture}
\end{center}
\vspace{-.25cm}
	\caption{Directed acyclic graph showing interoperability of the debiasing model output with a stochastic SIR epidemic model. Latent nodes for \gls*{ltla} $\ltlasub$'s prevalence and $\Rt$ number in week $\weeksub$,  $\prevprop_\subiw$ and $\Rt_\subiw$ are shown, along with weekly $\pill$ test data, integrated via the debiasing model-outputted marginal likelihood at \eqref{eq:marg_likelihoods}. Details of the discrete time Markov chains on the latent nodes are given in equations \eqref{eq:pi_markov}-\eqref{eq:recov_markov}.}
	\label{fig:sir_graphical_model}
\end{figure}

The marginal likelihood $p(\text{$\nt_\subiw$\ of\ $\Nt_\subiw$} \mid \prevprop_\subiw)$ from \eqref{eq:marg_likelihoods} can be used to link the 
$\pill$ data directly to latent prevalence nodes in a graphical model. As a concrete example, we implemented a full Bayesian version of the standard stochastic SIR model \cite{brauer_fred_mathematical_2008,nicholson_local_2021,scott_bayesian_2002}.
Figure~\ref{fig:sir_graphical_model} illustrates the SIR model DAG, relating 
prevalence proportion $\prevprop_\subiw$, effective reproduction number $\Rt_{\ltlasub,\weeksub}$, and test data $(\pilln)_{\subiw}\equiv \pillniw$. Note that Figure~\ref{fig:sir_graphical_model} is a special case of Figure~\ref{fig:marginal_approx_lik_graph}(b) with data node $Y$ empty and $\theta = \Rt_{\ltlasub, 1:W}$. In Figure~\ref{fig:sir_graphical_model} each $\prevprop_\subiw$ is linked to its corresponding test data $(\pilln)_{\subiw}$ via the
marginal likelihood $p(\pillniw \mid \prevprop_\subiw)$. Consecutive $\prevprop_\subiw$ and $\Rt_\subiw$ nodes are related by a discrete time Markov chain, in which the stochastic change in the number infected at week $\weeksub$, relative to week $\weeksub -1$, is modelled as the difference between a Poisson-distributed number of new infections and a Binomial-distributed number of new recoveries:
\begin{align}
\label{eq:pi_markov}
    \prevprop_\subiw \popsize_\ltlasub - \prevprop_{\ltlasub,\weeksub - 1} \popsize_\ltlasub \mid \prevprop_{\ltlasub,\weeksub - 1}, \Rt_{\ltlasub,\weeksub - 1} & = \text{\# new infections} - \text{\# recoveries}\\
\label{eq:infect_markov}
    \text{\# new infections} \mid \prevprop_{\ltlasub,\weeksub - 1}, \Rt_{\ltlasub,\weeksub - 1} &\sim \text{Poisson}(\gamma \Rt_{\ltlasub,\weeksub - 1} \prevprop_{\ltlasub,\weeksub - 1} \popsize_\ltlasub) \\
\label{eq:recov_markov}
    \text{\# new recoveries} \mid \prevprop_{\ltlasub,\weeksub - 1}&\sim \text{Binomial}(\prevprop_{\ltlasub,\weeksub - 1} \popsize_\ltlasub,\  \gamma)\ ;
\end{align}
$\gamma$ is the (pre-specified) probability of recovery from one week to the next; and the $\Rt_{\ltlasub,\weeksub}$ are modelled sequentially by an AR1 process.
We sample from the full Bayesian posterior for this SIR model using MCMC methods (see \cite{nicholson_local_2021} for details).

\begin{figure}[h]
	\begin{adjustbox}{max totalsize={\textwidth}{\textheight},center}
	    \includegraphics{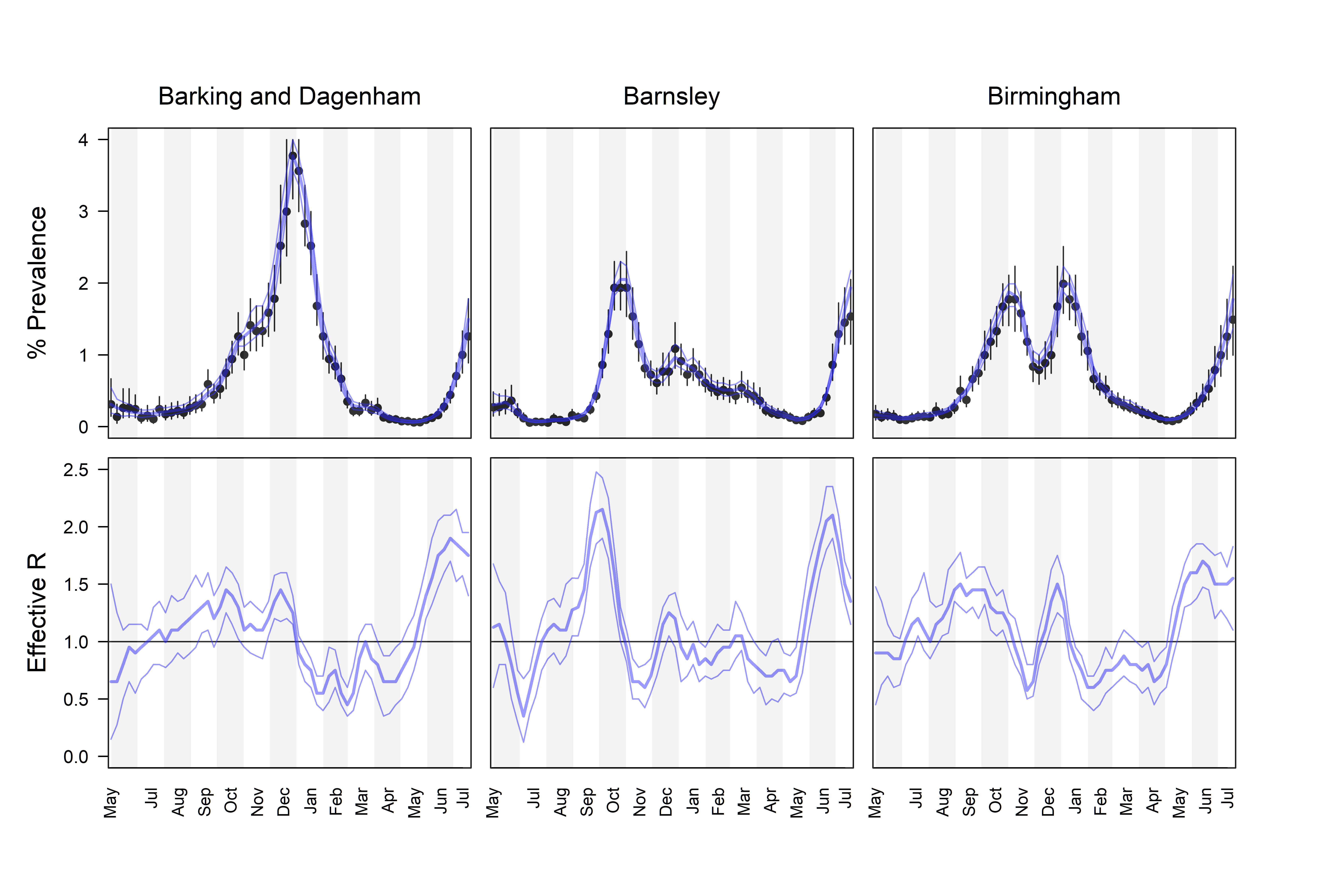}
	\end{adjustbox}
	\caption{SIR longitudinal posterior (\% prevalence and $\Rt$) compared with cross-sectional posterior for \% prevalence. Top panels: cross-sectional \% prevalence posterior median and 95\% CIs (black points and whiskers), and SIR modelled \% prevalence posterior median and 95\% CIs (blue curves). Bottom panels: SIR modelled longitudinal $\Rt_\subiw$ posterior median and 95\% CIs.}\label{fig:sir_interop}
\end{figure}

Figure~\ref{fig:sir_interop} compares, for three example \glspl*{ltla}, the cross-sectional posteriors for $\prevprop$ with the SIR-model MCMC-sampled longitudinal posteriors for $\prevprop$ and $\Rt$. The width of the SIR posterior 95\% credible intervals are often much narrower than the cross-sectional posterior CI width, illustrating the benefit of sharing prevalence information across time points within the framework of an epidemiological model. Fitting the full Bayesian SIR model provides posterior credible intervals on the effective reproduction number $\Rt$ (Figure~\ref{fig:sir_interop} bottom panels), which is an important measure of spatiotemporally local rates of transmission. In Section~\ref{sec:epimap_modelling}, we compare these estimates of local $\Rt$ (based on $\pill$ from a single LTLA only) with spatially smoothed estimates of local $\Rt$ from Epimap \cite{teh2021}.

\subsection{Interoperability between debiasing and space-time equality analysis}
\label{sec:st_modelling}


There is extensive evidence to suggest that ethnically diverse and deprived communities have been differentially affected by the \gls*{covid} pandemic in the UK \cite{morales:2021, mathur:2021, rose:2020}. It thus is very important to be able to relate unbiased prevalence estimates such as those introduced in Section \ref{sec:debias} to area level covariates, in order to assess associations between the spread of the virus and particular population characteristics. Given the infectious nature of the disease, residual heterogeneity is likely to have a spatio-temporal structure, which needs to be accounted for in the model. Failing to account for autocorrelation may result in narrower credible intervals for parameters of interest and may inflate covariates effects \cite{lee2015controlling}. One option to incorporate sources of spatial autocorrelation in estimates of disease surveillance metrics such as prevalence, would be to run the debiasing model presented in Section~\ref{sec:cross_sectional_local_prevalence} augmented by  a spatio-temporal prior structure both on the $\prevprop_\subiw$ and on the ascertainment bias $\delta_\ltlasub$, while simultaneously adding the population characteristics as covariates to assess their impact on the spread of the disease. However, such a strategy would entail a prohibitive computational cost, not only due to the large number of latent variables typically involved in the specification of even a relatively simple spatio-temporal model, but also because of implementing a cut model in such a high-dimensional setting. As an alternative, in this section we illustrate two interoperable models which differ in how they treat the uncertainty of prevalence estimates. 

We focus on ethnic composition and socio-economic deprivation as covariates of interest, to assess the impact of socio-economic factors on the evolution of the pandemic. 
As our aim is to assess a link between prevalence proportion and the these two factors, first we specify a probability distribution for the outcome, $\prevprop_\subiw$ as:
\begin{equation}
   \text{logit}({\prevprop}_\subiw)| \eta_\subiw, \sigma^2 \sim \text{Normal}( \eta_\subiw, \sigma^2) 
 \end{equation}
where $\eta_\subiw$ is a linear predictor containing the area level variables of interest as well as space and time structured random effects, while $\sigma^2$ is the variance of the error term.

Since ${\prevprop}_\subiw$ is unknown, it is not possible to fit this model directly. A first interoperable approach, which we call the \textit{naive model}, simply consists in plugging-in the debiased prevalence proportion point estimates (e.g. median) $\widehat{\prevprop}_\subiw$, outputted by the debiasing model of Section~\ref{sec:debias}:
\begin{equation}\label{eq:naive}
   \text{logit}(\widehat{\prevprop}_\subiw)| \eta_\subiw, \sigma^2 \sim \text{Normal}( \eta_\subiw, \sigma^2) 
 \end{equation} 
This model considers the prevalence estimates as fixed quantities, neglecting their uncertainty. 

A second interoperable approach accounts for the posterior uncertainty from the debiased prevalence model via \eqref{eq:gaussmomlik}, similarly to \cite{Pirani:2020, dominici2000combining}. In practice, for the $i$-th LTLA and $w$-th week this \textit{heteroscedastic model} reformulates (\ref{eq:naive}) to include the variance component $\hat{\sigma}^2_\subiw$:

\begin{equation}
   \text{logit}(\widehat{\prevprop}_\subiw)| \eta_\subiw, \sigma^2 \sim \text{Normal}( \eta_\subiw, \hat{\sigma}^2_\subiw + \sigma^2)  
 \end{equation} 

In both models, we use the following specification for the linear predictor $\eta_\subiw$, in order to assess the effect of the area level variables of interest (proportion of BAME and IMD score) on the prevalence estimates: 
\begin{equation}
  \eta_\subiw = \beta_0 + \beta_1 {\tt BAME}_{i} + \beta_2 {\tt IMD}_i + \lambda_{i} + \epsilon_{w}  
  \label{eq:log_reg_tp}
\end{equation}
where $\beta_0$ is the global intercept $\{\beta_1,\beta_2\}$ quantify the effects of the covariates of interest, $\lambda_i$ denotes the area specific random effect accounting for spatial autocorrelation, and $\epsilon_w$ represents the temporal random effect. Details on the model specification can be found in Appendix \ref{app:st-model}.

\begin{figure}
    \centering
    \includegraphics[width = 0.95\textwidth]{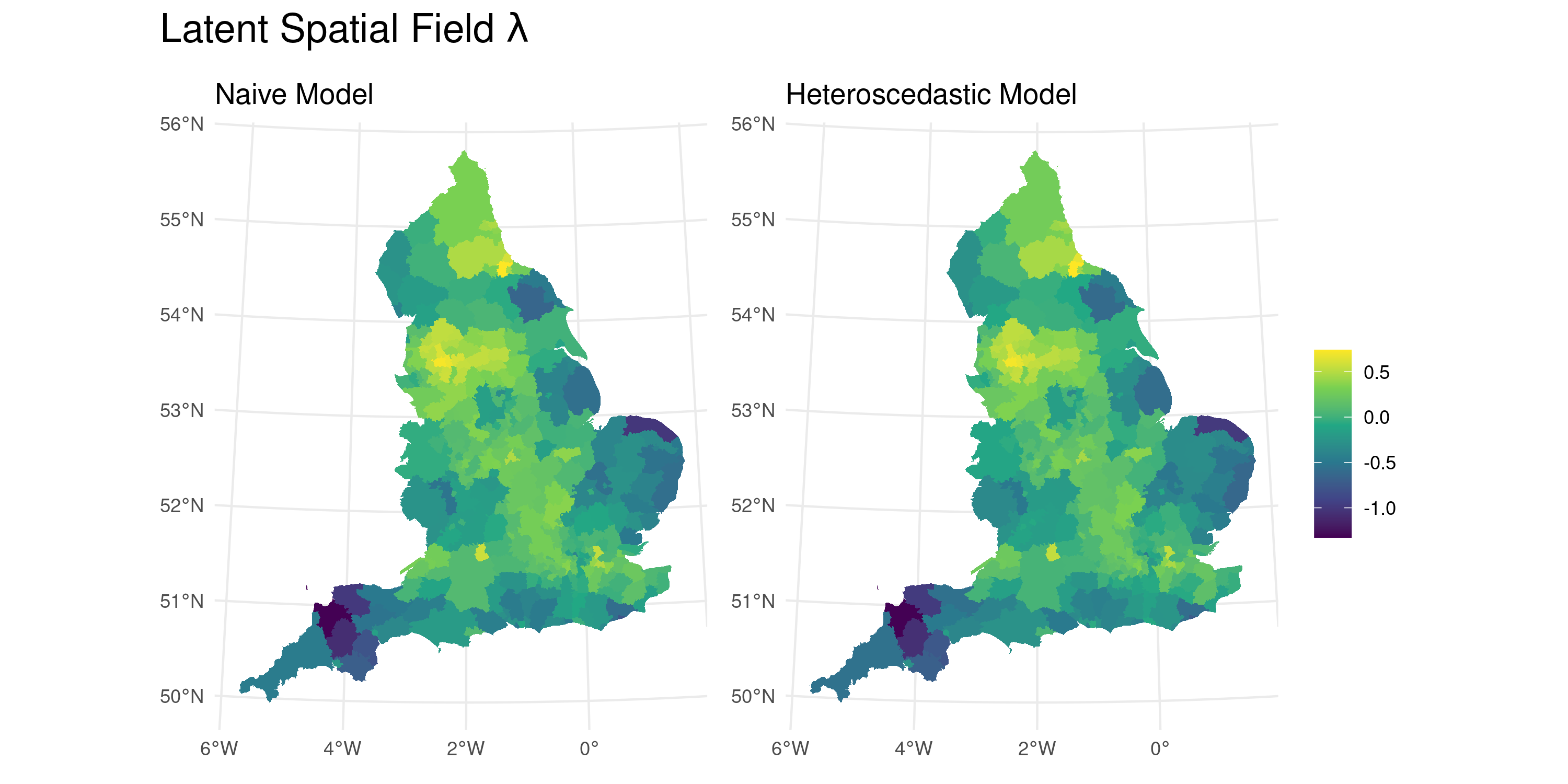}
    \caption{Posterior median of the spatial random effect $\bm{\lambda}$}
    \label{fig:spatial_res}
\end{figure}
\begin{figure}
    \centering
    \includegraphics[width = 0.95\textwidth]{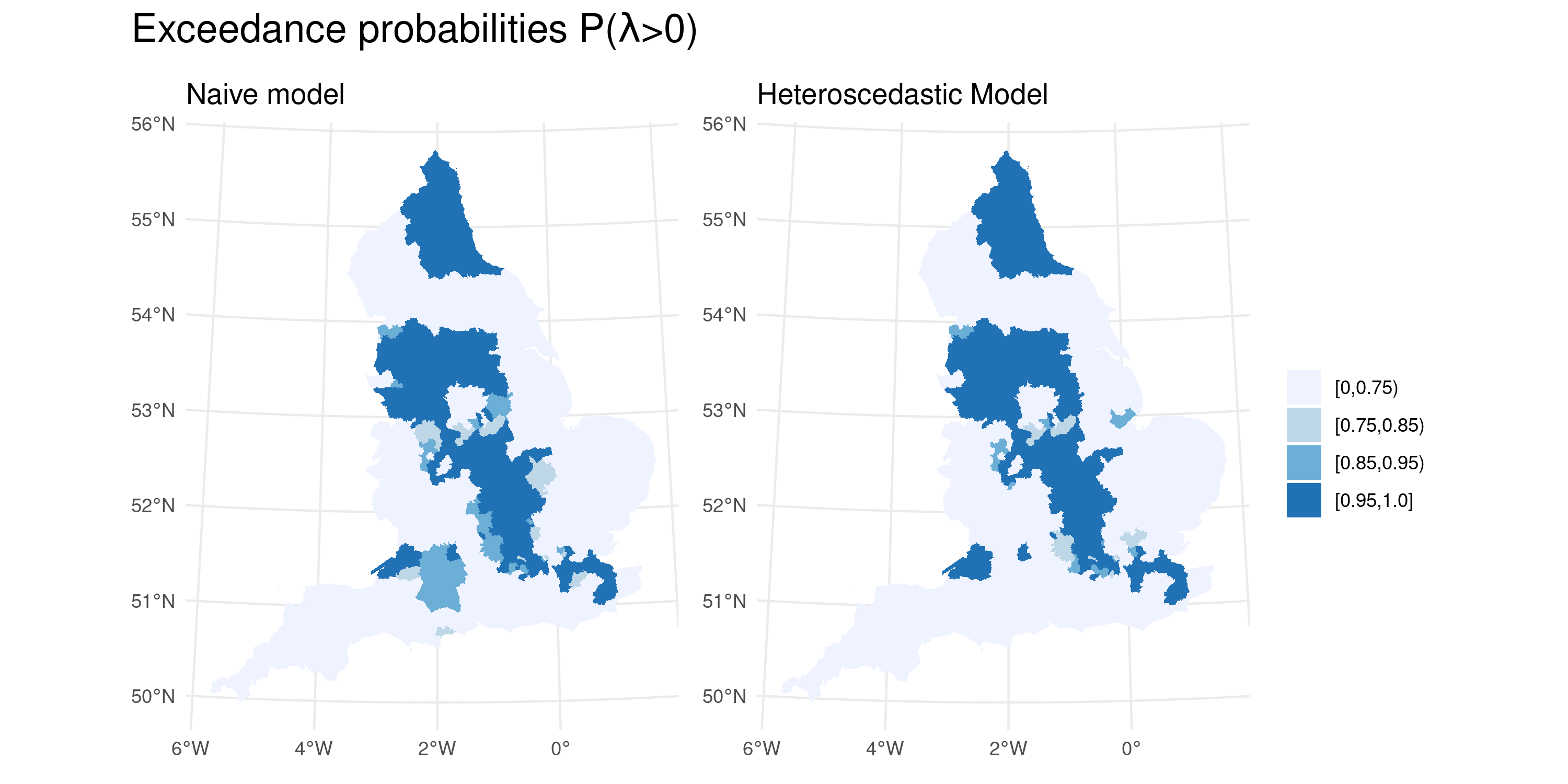}
    \caption{Posterior probability of having a positive spatial residual.}
    \label{fig:spatial_exc}
\end{figure}

Output from the two interoperable modelling approaches presented above is used to examine the effects of socio-economic factors (i.e. ethnic diversity and socio-economic deprivation) on \gls*{covid} unbiased prevalence. 
Additionally, the models allow us to characterise the baseline spatial distribution of prevalence across LTLAs in England, \textit{after accounting for their socio-economic and ethnic profiles}. This enables us, for example, to identify LTLAs that have been particularly badly affected by \gls*{covid} relative to their level of deprivation and size of BAME population and as a result may require further consideration by policy-makers.

While both models are examples of interoperability, they differ in how they handle uncertainty in the debiased prevalence proportion outcome measure; the naive model treats the prevalence estimates as fixed quantities, while the heteroscedastic model propagates their uncertainty through an additional variance term. This difference is predominantly reflected in the uncertainty of parameter estimates. Although the posterior median point estimates for the effects of IMD and BAME (Table~\ref{tab:equality}) and the underlying spatial fields (Figure~\ref{fig:spatial_res}) are largely indistinguishable, the precision for the global error term (i.e., $1/\sigma^2$) is twice as large in the heteroscedastic model. Precision for the spatial component is also larger in the heteroscedastic model compared to the naive model, suggesting that the global error and the spatial random effect in the naive model were capturing part of the spatial variability present in the prevalence proportion estimates. It is also interesting to note that the exceedance probability map corresponding to the heteroscedastic model is sharper, showing that the ranking of areas deviating from the national average is influenced by the inclusion of uncertainty on the debiased prevalence estimates (Figure~\ref{fig:spatial_exc}).

 \begin{table}[h]
 \centering
 \caption{Posterior mean and corresponding 95\% Credible Interval for the parameter estimates of the Naive (left) and Heteroscedastic (right) space-time equality analyses. Estimates for the fixed effect coefficients are on the Odds Ratio scale. The precision of the time and spatial random effects, $1/\sigma^2_{\epsilon}$ and $\tau$, are defined in Appendix~\ref{app:st-model}.}
 \begin{tabular}{l rrrr}
 & \multicolumn{2}{c}{Naive} &  \multicolumn{2}{c}{Heteroscedastic} \\
   \hline
        & Median & 95\% CI &  Median & 95\% CI \\ 
   \hline
    BAME effect ($\beta_1$) & 1.21 & (1.16, 1.27)  & 1.20 & (1.15, 1.25) \\ 
   IMD effect ($\beta_2$)  & 1.12 & (1.08, 1.16)  & 1.11 & (1.07, 1.15)  \\ 
    \hline
Precision of the Gaussian residuals ($1 / \sigma^2$) & 2.61 & (2.56, 2.67) & 4.02 & (3.92, 4.12) \\ 
Precision of time random effect ($1/\sigma^2_{\epsilon}$) & 31.00 & (18.94, 47.82) & 30.66 & (19.62, 47.38) \\ 
Precision of spatial random effect ($\tau$) & 9.83 & (8.15, 11.91) & 11.79 & (9.88, 14.55) \\ 
       \hline

 \end{tabular}
 \label{tab:equality}
 \end{table}


\subsection{Interoperability between the debiasing and Epimap models}
\label{sec:epimap_modelling}

Epimap is a hierarchical Bayesian method for estimating the local instantaneous reproduction number $\Rt_\subit$ that models both temporal and spatial dependence in transmission rates \cite{teh2021}. Epimap incorporates information on population flows to model transmission between local regions, and performs spatiotemporal smoothing on the $\Rt_\subit$. The data inputted to Epimap are daily positive $\pill$ test counts at \gls*{ltla} level,  the $\nt_\subit$ in our notation.

\subsubsection{\textbf{Epimap model overview.}}\label{sec:epimap_overview} The observation model for $\nt_\subit$ is an overdispersed negative binomial model with mean $E_{\ltlasub,\daysub}$ given by a convolution of a testing delay distribution $\dtest_s$ with the past incidences $X_{\ltlasub,1:t}$ \cite{flaxman2020estimating}, i.e.\ 
\begin{align}
   \label{eq:epimap_obs_model1}
     \nt_{\ltlasub,\daysub} \mid X_{\ltlasub,1:\daysub} &\sim \text{NegBin}(E_{\ltlasub,\daysub},\ \phi_\ltlasub)\\ 
       \label{eq:epimap_obs_model2}
     E_{\ltlasub,\daysub} &\equiv\sum_{s=0}^\daysub X_{\ltlasub,\daysub-s } \dtest_s
\end{align}
where  $\phi_\ltlasub \sim \mathcal{N}_+(0,5)$. 

\gls*{ltla} $i$'s incidence $X_\subit$ is probabilistically modelled conditional on past incidences $X_{1:n,1:t-1}$ via local and cross-coupled infection loads, denoted $Z_\subit$ and $\tilde{Z}_\subit$ respectively. Specifically, the local infection load $Z_\subit$ is given by a convolution of $W_s$ with the past incidences $X_{\ltlasub,1:t}$, where the generation distribution $W_s$ is the probability that a given transmission event occurs $s$ days after the primary infection. The local infection loads contribute transmission events not only locally but also to other regions, giving the cross-coupled infection load $\tilde{Z}_\subit$, with inter-regional transmission defined via a flux matrix $F$, built on the commuter flow data introduced in Section~\ref{sec:commuter_flow}, in which $F_{ji}$ denotes the probability that a primary case based in area $j$ infects a secondary case based in area $i$. In summary, the local and cross-coupled infection loads are defined as
\begin{align}
\label{eq:epimap_infection_loads}
      Z_\subit\equiv \sum_{s=1}^t X_{\ltlasub, t-s}W_s &\hspace{1.5cm} \tilde{Z}_\subit\equiv \sum_{j=1}^n F_{ji}^{(t)} Z_{j,t}
  \ .
\end{align}

The incidence $X_\subit$ follows an overdispersed negative binomial distribution with mean given by the product of the reproduction number $\Rt_\subit$ with the cross-coupled infection load $\tilde{Z}_\subit$:
\begin{align}
\label{eq:incidence_distribution}
    X_\subit \mid \Rt_\subit,X_{1:n,1:t-1} &\sim \text{NegBin}(\Rt_\subit \tilde{Z}_\subit,\ \phi)\ .
\end{align}
Note that the $\Rt_\subit$ are our primary inferential target, and are estimated via the posterior distribution of the ratio $X_\subit/\tilde{Z}_\subit$. In Figures~\ref{fig:epimap_maps} and \ref{fig:epimap_R_counts_comp} below we present the posterior distribution of the weekly averaged ratio, $\Rt_\subiw := \frac{1}{7}\sum_{t\in w} X_\subit/\tilde{Z}_\subit$. A final aspect of the Epimap model is the smoothing on $\Rt_\subit$, whereby information is shared across time and space through specification of various Gaussian processes on $\log \Rt_\subit$ (see \cite{teh2021} for details).

\subsubsection{\textbf{Probabilistic interface between Epimap $\Emod$ and debiasing model $\Dmod$.}} \label{sec:epimap_interface} There are three immediate and important differences between Epimap and the debiasing model that require attention at the model interface. 
First, Epimap's measure of infection burden is incidence (i.e.\ the number of new infections contracted in a time interval), whilst the debiasing model is based on point prevalence, as we defined in Section~\ref{sec:modelling}.
Second, Epimap is at daily frequency, indexed $\daysub$, while the debiasing model is at weekly frequency, indexed $\weeksub$. Third, some LTLAs were recently merged to create a more coarse-scale local geography, and the debiased model works with the newer coarser LTLA geography, while Epimap still works with the older finer-scale LTLA geography. 

To map from incidence to prevalence, we draw from the existing \gls*{covid} literature on the probability of testing \gls*{pcr} positive when swabbed $s$ days post infection \cite{hellewell_estimating_2020}; we denote this as
\begin{align}
\label{eq:define_pi_weekly}
    \dpcr_s &: =\Pp(\text{would\ test\ }\wtpcr \text{\ on\ day\ }s \mid \text{contract\ virus\ day\ 0})\ .
\end{align}
To address the daily-to-weekly mapping, we average the daily prevalence proportion (mapped from daily incidence) across the days of each week. To address the models' differences in LTLA geography, we are able straightforwardly to preserve the geographies of both models by deterministically aggregating Epimap's latent incidences as part of the mapping between models described below in \eqref{eq:daily_to_weekly} (details omitted for simplicity), thereby creating a seamless interface between models. In summary, the mapping from daily incidence $X_\subit$ to weekly prevalence proportion $\prevprop_\subiw$ is:
\begin{align}
    \label{eq:daily_to_weekly}
    \prevprop_\subiw &\equiv \frac{1}{7}\sum_{\daysub \in \weeksub} \prevprop_\subit\hspace{1.5cm}  \prevprop_\subit \equiv \frac{1}{\popsize_\ltlasub}\sum_{s=0}^\daysub X_{\ltlasub,\daysub - s} \dpcr_s\ ,
\end{align}
with $\popsize_\ltlasub$ denoting local population size. Finally, the latent weekly prevalence proportions $\prevprop_\subiw$ in the interoperable Epimap model are related to the debiasing model outputs $(\hat{\prevprop}_{\ltlasub,\weeksub}, \hat{\sigma}^2_{\ltlasub,\weeksub})$ via the approximate $\prevprop$-marginal likelihood of \eqref{eq:gaussmomlik} (see also Figure~\ref{fig:marginal_approx_lik_graph}(b)):
\begin{align}
    \label{eq:normal_lik_on_logit_scale}
    \text{logit}\left(\hat{\prevprop}_{\ltlasub,\weeksub}\right) \mid  X_{\ltlasub,1:\daysub[\weeksub]} &\sim \text{Normal}\left(\text{logit} \left(\prevprop_{\ltlasub,\weeksub}\right) , \hat{\sigma}^2_{\ltlasub,\weeksub}\right)\ ,
\end{align}
where $\daysub[\weeksub]$ denotes the final day in week $\weeksub$. 

In summary, the interface between Epimap and debiasing models is created by removing Epimap's observation model of \eqref{eq:epimap_obs_model1} and \eqref{eq:epimap_obs_model2}, and replacing it with the debiasing-model outputted marginal likelihood defined at \eqref{eq:daily_to_weekly} and \eqref{eq:normal_lik_on_logit_scale}.

\subsubsection{\textbf{Comparing and contrasting estimated $\Rt_\subiw$ across models.}} 
\begin{figure}[h]
	\begin{adjustbox}{max totalsize={\textwidth}{\textheight},center}
	    \includegraphics{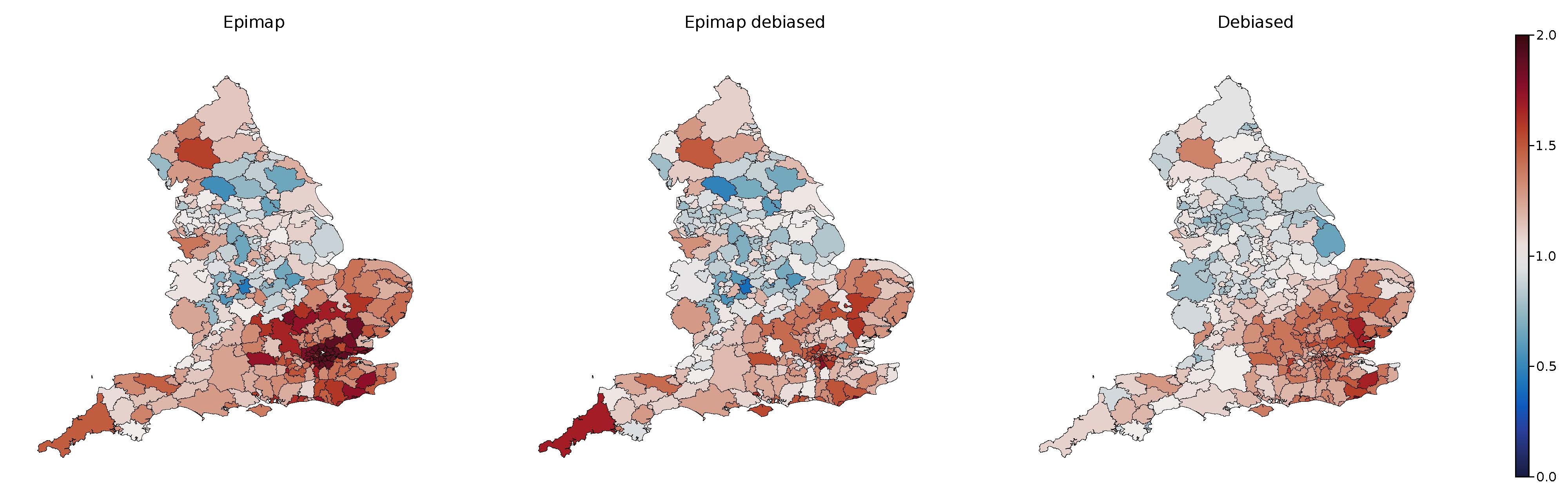}
	\end{adjustbox}
	\caption{Maps of estimated $\Rt_\subiw$ for three different models across all \glspl*{ltla} in England \gls*{wc} 4th December 2020. (a) Epimap ($\Emod$). (b) Epimap debiased interoperable model ($\EDmod$). (c) Debiased SIR model  ($\Dmod$).} \label{fig:epimap_maps}
\end{figure}

We estimated $\Rt_\subit$ under each of the three models: Epimap ($\Emod$) described in Section~\ref{sec:epimap_overview}; the debiased SIR model ($\Dmod$) output as described in Section~\ref{sec:SIR_debiased}; and the interoperable combination of Epimap and debiased models ($\EDmod$) described in Section~\ref{sec:epimap_interface}. Movie 1 in Supplementary Information provides a global perspective on the results, showing the longitudinal changes via maps evolving through time. 
Figure~\ref{fig:epimap_maps} shows one snapshot of this movie, for all \glspl*{ltla} in England \gls*{wc} 4th December 2020. 
The general theme is one of consistency in $\Rt_\subiw$ estimates across models, but there are points in space and time at which results differ. We will discuss these differences, demonstrating that they can help us to characterise and understand model performance. 

\subsubsection{\textbf{Interpreting model outputs via data synchronisation.}}
\label{sec:interpretation_via_data_sync} We turn first to the maps in Figure~\ref{fig:epimap_maps}, visually comparing and contrasting the three models across all \glspl*{ltla} in \gls*{wc} 4th December 2020. One interesting feature here is the presence of a few \glspl*{ltla} with low $\Rt_\subiw$ in both $\Emod$ and $\EDmod$, but with relatively high $\Rt_\subiw$ in $\Dmod$. The two most prominent, coloured blue in Figure~\ref{fig:epimap_maps}(a-b), are (North Warwickshire, Craven), which have $\Rt_\subiw$ estimated to be (\mbox{0.38
\unskip}, \mbox{0.47
\unskip}) by $\Emod$ and (\mbox{0.33
\unskip}, \mbox{0.41
\unskip}) by $\EDmod$, but estimated to be higher at (\mbox{0.85
\unskip}, \mbox{0.90
\unskip}) by $\Dmod$. When we compare two models that differ in both data inputs and in probabilistic structure (as do $\Emod$ and $\Dmod$) any difference in results cannot immediately be attributed solely to either data or model structure. However, by constraining the two models to have the same data inputs -- as we have here by using the prevalence outputs of $\Dmod$ as inputs to $\EDmod$ -- we can potentially learn more. Observing that $\EDmod$ agrees 
with its ``model twin'' $\Emod$, but disagrees with its ``data twin'' $\Dmod$, 
leads us to conclude that differing results in (North Warwickshire, Craven) between $\Dmod$ and $\Emod$ arise because of differences in model structure rather than because of differences in data inputs.

We examine the hypothesis that the differences in (North Warwickshire, Craven) $\Rt_\subiw$ between $\Emod$ and $\Dmod$ are attributable to Epimap's cross-coupled infection load, $\tilde{Z}_\subit$ in \eqref{eq:epimap_infection_loads}, which allows transmission across regional boundaries; in contrast, the debiased SIR model has only within-\gls*{ltla} transmission. Note from \eqref{eq:incidence_distribution} that Epimap's expected number of new infections is represented as the product $\mathbb{E} X_\subit =\Rt_\subit \tilde{Z}_\subit$, so that low estimates of $\Rt_\subit$ will arise when the cross-coupled infection load $\tilde{Z}_\subit$ is large relative to the latent incidence $X_\subit$. For 4th December 2020 in (North Warwickshire, Craven) $\EDmod$ outputs a posterior median for $X_\subit$ of (34.1, 27.7) and for $\tilde{Z}_\subit$ of (114.4, 81.1), 
consistent with the low $\Rt_\subit$ estimates of (\mbox{0.33
\unskip}, \mbox{0.41
\unskip}). We can further decompose the cross-coupled infection load $\tilde{Z}_\subit$ into infection load $Z_\subit$ originating from within \gls*{ltla} $i$ 
(27.0, 31.1), 
and $Z_{-\ltlasub,t}$ originating from other \glspl*{ltla} 
(85.8, 49.9). 
See Figure~\ref{fig:epimap_external_infection_load} in Appendix~\ref{app:epimap_infection_load_figure} for a map of the proportion of infection load arising external to each \gls*{ltla}. It is clear that for (North Warwickshire, Craven) the majority of the infection load in the $\Emod$ (and $\EDmod$) models is \emph{external}, and that these are among only a handful of \glspl*{ltla} having external load at $>50\%$ of the infection burden. Through data synchronisation, theorizing on salient differences between models, and examining confirmatory diagnostic plots, we have increased our understanding of the operational differences between the $\Emod$ and $\Dmod$ models.

\subsubsection{\textbf{Illustrating synergy between models.}}
\begin{figure}[h]
	\begin{adjustbox}{max totalsize={\textwidth}{\textheight},center}
	    \includegraphics{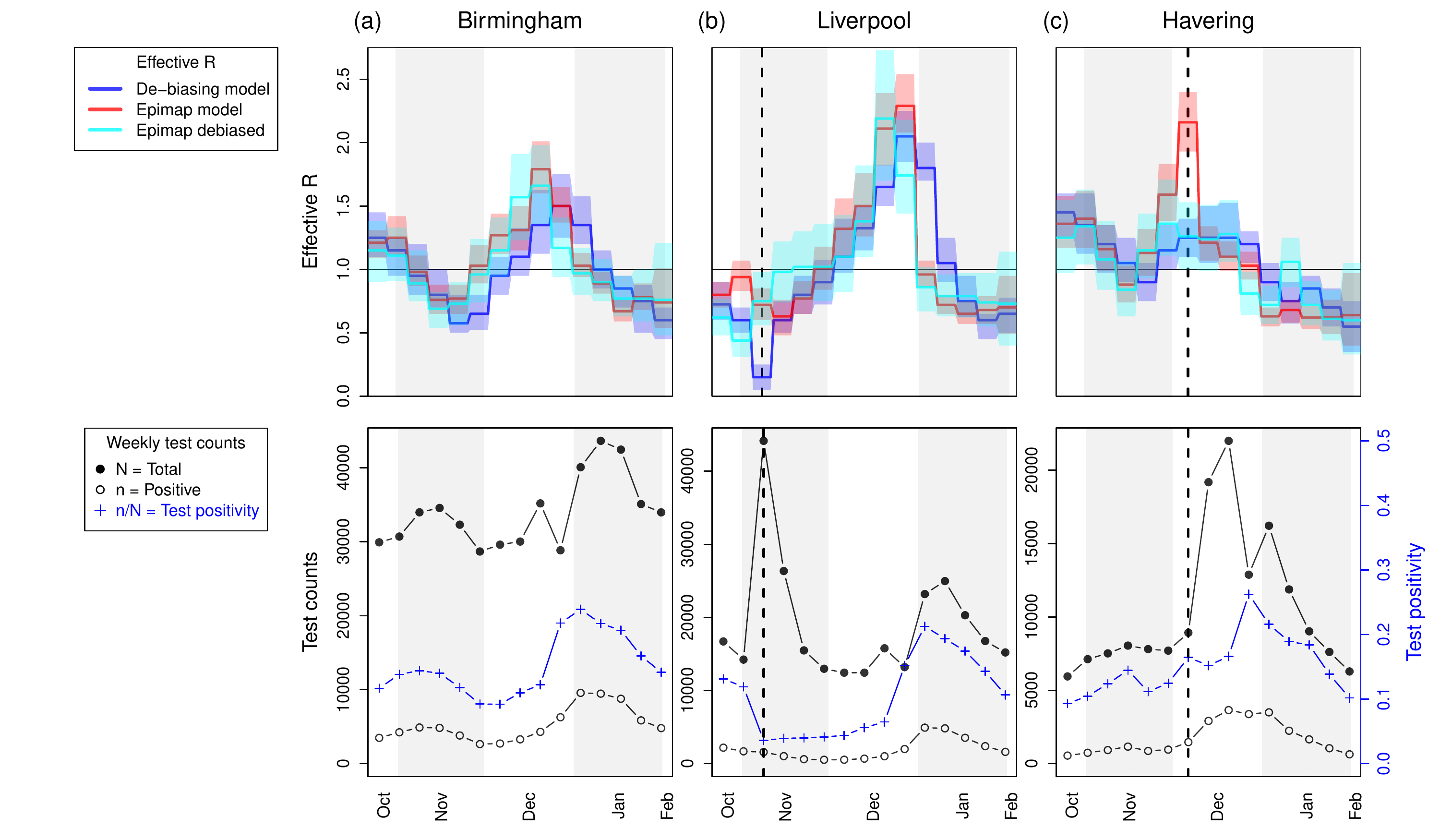}
	\end{adjustbox}
	\caption{Longitudinal $\Rt_\subiw$ trajectories superimposed above corresponding $\pill$ test data. Top: posterior median and 95\% CIs of $\Rt_\subiw$ for three models (see upper legend). Bottom: weekly test counts, positive $\nt_\subiw$ and total $\Nt_\subiw$, as well as test positivity $\nt_\subiw/ \Nt_\subiw$ (see lower legend). Vertical dashed lines in panels (b-c) indicate instances of model discordance in $\Rt_\subiw$ estimates, preceding or coinciding with local surges in testing capacity.}\label{fig:epimap_R_counts_comp}
\end{figure}

The top panels in Figure~\ref{fig:epimap_R_counts_comp} present longitudinal $\Rt_\subiw$ curves for four selected \glspl*{ltla} for each of the three models; the bottom panels display the corresponding $\pill$ weekly test counts and positivity rate. First note that the $\Rt_\subiw$ plot for Birmingham in Figure~\ref{fig:epimap_R_counts_comp}(a) exhibits reassuring similarity between models; indeed this is what we observe for the majority of \glspl*{ltla} that are not shown in the figure. But the \glspl*{ltla} in  Figure~\ref{fig:epimap_R_counts_comp}(b-c) display some interesting and contrasting behaviour between models.

The vertical dashed line in each panel \ref{fig:epimap_R_counts_comp}(b-c) coincides with, or immediately precedes, a surge in community testing capacity -- 
see the sharp increase in total test counts in the bottom panels at or after each dashed line. 
Figure~\ref{fig:epimap_R_counts_comp}(b) includes a mass testing pilot study in Liverpool beginning 6th November 2020 which led test positivity to drop sharply from \mbox{11.9
\unskip}\% in \gls*{wc} 29th October 2020 to \mbox{3.6
\unskip}\% the following week.
Such an abrupt and localized change in testing ascertainment is at odds with the spatiotemporally smooth parameterization of the debiasing model, leading to artefactual deflation of $\Rt_\subiw$ in $\Dmod$ in \gls*{wc} 5th November (marked with a dashed line). The $\Rt_\subiw$ deflation is not however evident in $\Emod$ or $\EDmod$. 

Turning to Figure~\ref{fig:epimap_R_counts_comp}(c) we note that, in mid-December 2020, parts of Essex and London, including the illustrated example of Havering, were moved into Tier 3 -- the very high alert level -- and earmarked for extra community testing. This led to a large spike in testing capacity and uptake, but test positivity rates remained relatively constant in the two weeks following the dashed line (in contrast to the mass testing pilot in Liverpool a month earlier: see bottom panels of  Figure~\ref{fig:epimap_R_counts_comp}(b-c)). This caused artefactual inflation of $\Rt_\subit$ in $\Emod$, since positive cases surged in line with testing capacity, and $\Emod$ takes as input only the positive cases $\nt_\subit$ (but not the total tests $\Nt_\subit$). The other two models, $\EDmod$ and $\Dmod$, are not obviously affected by the December 2020 testing surge in Figure~\ref{fig:epimap_R_counts_comp}(c). 

The steady performance of the interoperable model $\EDmod$ across Figure~\ref{fig:epimap_R_counts_comp} points to a desirable synergistic form of robustness -- we aspire to synthesise models in such a way that they support one another, with the strengths of one model stabilising inference if and when the other model shows any weakness with respect to the data generating mechanism. The suboptimal $\Rt_\subiw$ estimation observed in \ref{fig:epimap_R_counts_comp}(b-c) occurred when there were sudden changes in the ascertainment mechanism; it is reassuring that both adversely affected models ($\Dmod$ in Figure~\ref{fig:epimap_R_counts_comp}(b) and $\Emod$ in \ref{fig:epimap_R_counts_comp}(c)) apparently return to agreement with the other models just one week after these extreme shifts in ascertainment bias.

\section{Discussion}
Based on our experience, we believe that striving for interoperability across all facets of the delivery of statistical projects will provide:
\begin{itemize}
    \item \emph{Agility:} the ability rapidly to interlink and recycle statistical modelling outputs across analyses, with components transferable across health security problems;
    \item \emph{Robustness:} the structural assembly of modules that can be tested independently and connected in such a way as to mitigate any widespread impact of model misspecification; 
    \item \emph{Sustainability:} a shareable, high-quality, reusable, open-source analytic code base of modules that grows over time;
    \item \emph{Transferability:} a way to facilitate co-ownership of projects with public health and health policy teams, allowing rapid impact from academia and industry to be delivered against relevant, time-sensitive problems;
    \item \emph{Preparedness:} solutions built for a specific health emergency, such as the \gls*{covid} pandemic, can be re-purposed to meet future public health challenges. In particular, the necessary generic structural links between the data engineering architecture and the analytic and modelling side will have been already built. 
\end{itemize}

Many challenges lie ahead on the path towards an effective, interoperable, and comprehensive disease surveillance system. From a statistical point of view, it is most relevant to focus our attention on  issues that are generic and likely to recur when addressing a range of questions. For brevity we will only mention three particularly challenging ones. 

A major hurdle that interoperability will face is the need to integrate evidence from data collected at different time steps and spatial scales. For example, the time granularity of the randomised survey data used in our debiased prevalence model is the week, yet most epidemic models have been built on the basis of daily case numbers, thus necessitating an additional time-alignment interface. Similarly, it will be common to have to integrate different geographies into a single model, constrained by the data sources. Misaligned geographies is a recurrent statistical issue that has been much discussed in environmental sciences \cite{mugglin_fully_2000}, and for which pragmatic but robust solutions need to be investigated. 

A second challenge is situations when moment-matched Gaussian distributions, as used here, do not provide adequate approximations; alternatives include particle-based approaches, in which posterior samples from a module are used as a proposal within an MCMC scheme \cite{lunn_twostage_2013, goudie_joining_2019} or for importance sampling \cite{mauff_joint_2020} or within a sequential Monte Carlo scheme \cite{lindsten_divideconquer_2016}. 

A final challenging issue that we have already encountered, and dealt with in Section \ref{sec:modularized} by using a cut posterior, is how best to balance or weight different sources of evidence, to take into account prior knowledge; see \cite{de_angelis_four_2015} for a discussion of evidence weighting from an epidemiological perspective. Rather than completely preventing feedback, as per a cut posterior, it may be desirable to only partially down-weight, as proposed by \cite{carmona_semi-modular_2020}. For example, in the case of diagnostic tests, weights might take into consideration their modus operandi and context of use.

Although the analyses presented here were motivated by the specific example of the \gls*{covid} pandemic, the overarching principle of interoperability may be pertinent in a variety of contexts with complex modelling requirements in a dynamic environment, such as climate change or natural disaster management.

\section*{Funding}

MB acknowledges partial support from the MRC Centre for Environment and Health, which is currently funded by the Medical Research Council (MR/S019669/1). RJBG was funded by the UKRI Medical Research Council (MRC) [programme code MC\_UU\_00002/2] and supported by the NIHR Cambridge Biomedical Research Centre [BRC-1215-20014]. BCLL was supported by the UK Engineering and Physical Sciences Research Council through the Bayes4Health programme [Grant number EP/R018561/1] and gratefully acknowledges funding from Jesus College, Oxford. SR is supported by MRC programme grant MC\_UU\_00002/10; The Alan Turing Institute grant: TU/B/000092;  EPSRC Bayes4Health programme grant: EP/R018561/1. HG and TF acknowledge partial support from Huawei Research UK. Infrastructure support for the Department of Epidemiology and Biostatistics is also provided by the NIHR Imperial BRC. Authors in The Alan Turing Institute and Royal Statistical Society Statistical Modelling and Machine Learning Laboratory gratefully acknowledge funding from the Joint Biosecurity Centre, a part of NHS Test and Trace within the Department for Health and Social Care. This work was funded by The Department for Health and Social Care with in-kind support from The Alan Turing Institute and The Royal Statistical Society. The computational aspects of this research were supported by the Wellcome Trust Core Award Grant Number 203141/Z/16/Z and the NIHR Oxford BRC. The views expressed are those of the authors and not necessarily those of the National Health Service, NIHR, Department of Health, Joint Biosecurity Centre, or PHE.

\begin{appendix}
\section{\textbf{Longitudinal smoothing prior for bias parameter $\biasparpill$.}} 
\label{sec:smoothing_prior_for_delta}

We evaluate the cross-sectional cut posterior for $\biasparpill_\subjw$, the bias at week $\weeksub$ in \gls*{phe} region $\phesub$ (Figure~\ref{fig:cut_model}(a)), and use these to construct a prior to take forward to full Bayesian inference at each \gls*{ltla} in region $\phesub$. To induce smoothness we construct a ``product-of-experts'' prior \cite{hinton_training_2002}: 
\begin{align}
\label{eq:delta_given_nu_mvprior}
     p(\B{\delta}_{\phesub,1:W}) &\propto \text{Normal}(\B{\delta}_{\phesub,1:W}  \mid \B{0}, \B{\Sigma}_\delta)\ \prod_{\weeksub\in \mathcal{W}} \mathrm{N}(\delta_\subjw \mid \hat{m}_\subjw, \hat{s}^2_\subjw)\prod_{\weeksub\not\in \mathcal{W}} \mathrm{N}(\delta_\subjw \mid 0, v_{\text{flat}})\ .
\end{align}
The first term on the right of \eqref{eq:delta_given_nu_mvprior} is a subjective prior on the longitudinal smoothness of $\biasparpill_{1:T}$ encoding an AR1 process in $\B{\Sigma}_\delta$, defined to be only very weakly informative with respect to average location of $\B{\delta}_{\phesub,1:W}$; the second term is the product of (approximations to) the cross-sectional cut-posterior marginals from \eqref{eq:cut_model} at weeks $\mathcal{W}$ for which \gls*{react} data are available, i.e.\ 
\begin{align}
    \prod_{\weeksub\in \mathcal{W}}p^{\text{cut}}(\biasparpill_\subjw \mid \reactnjw, \pillnjw) &\approx  \prod_{\weeksub\in \mathcal{W}} \mathrm{N}(\delta_\subjw \mid \hat{m}_\subjw, \hat{s}^2_\subjw)\ ;
\end{align}
and the third term is a product of noninformative cross-sectional priors when \gls*{react} data are unavailable. 

The normalised form of \eqref{eq:delta_given_nu_mvprior} is MV Gaussian: \begin{align}
\label{eq:delta_given_nu_mvprior_v2}
    p(\B{\delta}_{\phesub,1:W}) &= \mathrm{Normal} \biggl(\B{\delta}_{\phesub,1:W}\ \biggl|\  \B{\hat{\mu}}_{\biasparpill}^{\text{out}},\ \B{\hat{\Sigma}}_{\biasparpill}^{\text{out}}  \biggr)\\
    \nonumber
    \B{\hat{\mu}}_{\biasparpill}^{\text{out}}&:=(\B{\Sigma}_\delta^{-1} + \B{D}^{-1})^{-1} \B{D}^{-1}\B{\hat{\mu}} \\
    \nonumber
    \B{\hat{\Sigma}}_{\biasparpill}^{\text{out}}&:=- (\B{\Sigma}_\delta^{-1} + \B{D}^{-1})^{-1}
\end{align}
with ($\B{\hat{\mu}}$, diagonal matrix $\B{D}_{W\times W}$) having elements $(\hat{m}_\subjw, \hat{s}^2_\subjw)$ for $\weeksub\in \mathcal{W}$ and $(0, v_{\text{flat}})$ for $\weeksub\not\in \mathcal{W}$. 

We denote the marginal distribution of \eqref{eq:delta_given_nu_mvprior_v2} for week $\weeksub$ by  
\begin{align}
    p_{\text{\gls*{dd}}}(\delta_{\phesub,w})  &:= \mathrm{Normal}(\delta_{\phesub,w}\mid \hat{\mu}_\subjw, \hat{\tau}^2_\subjw)\ ,
\end{align}
deploying these $p_{\text{\gls*{dd}}}(\delta_{\phesub[\ltlasub],w})$ as data-dependent priors independently at each week $\weeksub$ in \gls*{ltla} $\ltlasub$ in region $\phesub[\ltlasub]$ as described in Section~\ref{sec:cross_sectional_local_prevalence}. 

\section{Full Model Specification for the Space-time Equality Analysis}\label{app:st-model}
Following \cite{simpson2017penalising, riebler2016intuitive}, we model $\bm{\lambda} = (\lambda_1, \dots, \lambda_{I})$, the random effect accounting for the spatial autocorrelation across LTLAs, as 
\begin{equation}
\bm{\lambda} = \frac{1}{\sqrt{\tau}} \left(\sqrt{1-\rho} \bm{v}+ \sqrt{\rho}\bm{u}\right).
\end{equation}
The vector $\bm{u} = (u_1, \dots, u_{I})$ is a spatially structured random effect with prior distribution 
\begin{equation}
    \bm{u}| \tau, \rho \sim \text{Normal}(\bm{0}, \bm{Q^-})
\end{equation} 
where $\bm{Q^-}$, is the inverse of the precision matrix of a ICAR model, scaled in the sense of \cite{sorbye2014scaling}. The vector $\bm{v} = (v_1, \dots, v_I)$ is an i.i.d. Gaussian random effect, that is 
\begin{equation}
    \bm{v}| \tau, \rho \sim \text{Normal}(\bm{0}, \bm{I})
\end{equation}
where $\bm{I}$ is the identity matrix. To account for the time dependence, $\epsilon_{w}$ is modelled as a random walk of order $2$. Given $\Delta^2\epsilon_w = \epsilon_w - 2\epsilon_{w+1} + \epsilon_{w+2}$, this can be formalized as
\begin{equation}\Delta^2\epsilon_w|\sigma^2_{\epsilon} \sim \text{Normal}\left(0, \sigma^2_{\epsilon} \right).\end{equation}
Both $\bm{\lambda}$ and $\bm{\epsilon}$ imply a degree of smoothing in space and in time and help highlight persistent patterns in the data.
\\*
Finally we set a non-informative Gamma$(1, 5\times10^{-5})$ prior on the inverse of $\sigma^2 _{\epsilon}$ and a non-informative Normal$(0, 1000)$ prior for the fixed effect coefficients $\beta_1$ and $\beta_2$. All models are fitted using the R package {\tt INLA} \cite{rue2009approximate}.

\section{Illustration of Epimap external infection load}\label{app:epimap_infection_load_figure}

\begin{figure}[h]
	\begin{adjustbox}{max totalsize={\textwidth}{.4\textheight},center}
	    \includegraphics{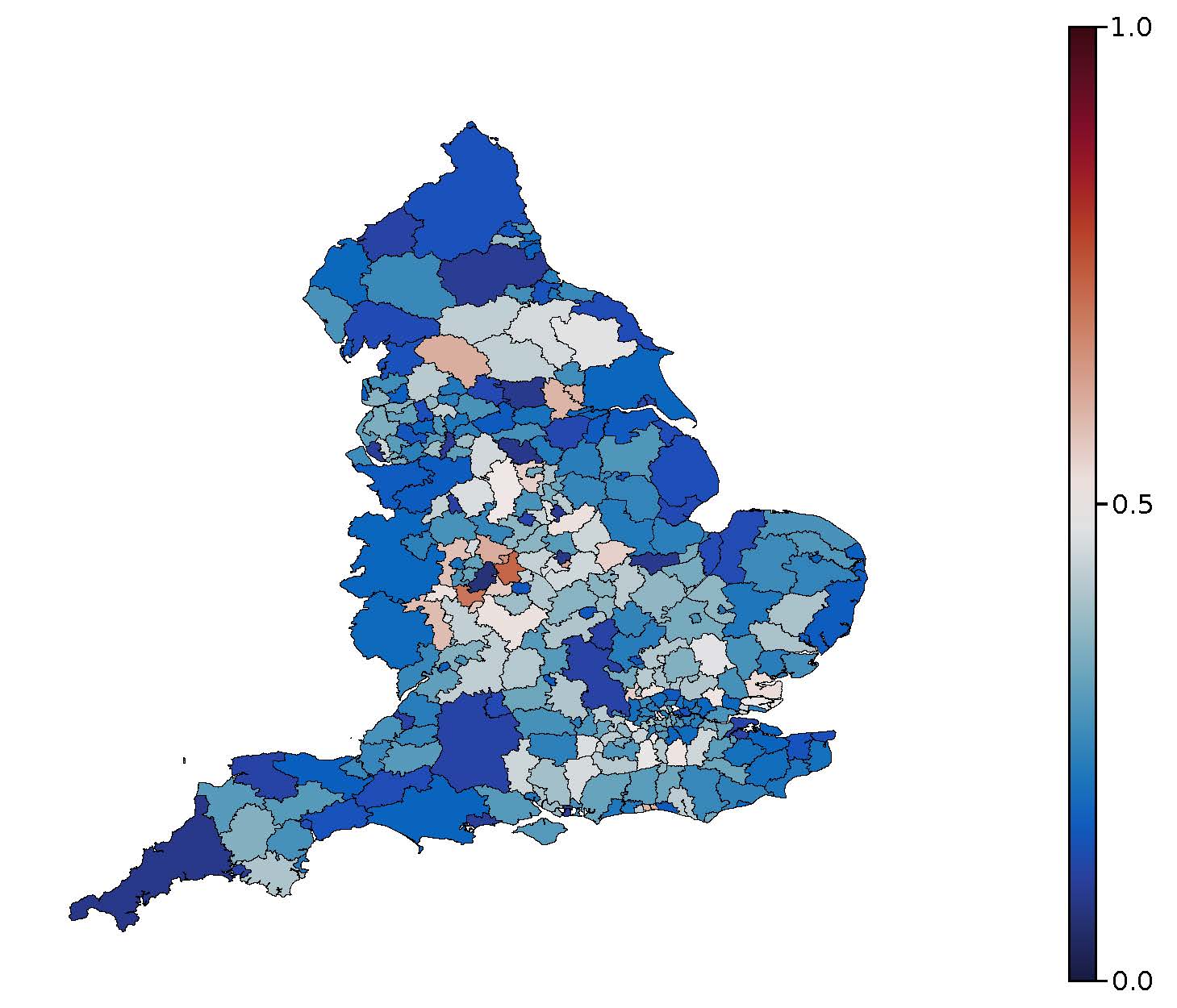}
	\end{adjustbox}
	\caption{Map of the proportion of infection load originating external to each LTLA on 4th December 2020. In Section~\ref{sec:interpretation_via_data_sync} we discuss the relatively low estimates of $\Rt_\subiw$ seen for (North Warwickshire, Craven), and we attribute this to them experiencing a relatively large external infection load, (North Warwickshire, Craven) are among the approximately 10 \glspl*{ltla} in this map with external infection load proportionally $> 50\%$. \label{fig:epimap_external_infection_load}}
\end{figure}

\end{appendix}

\newpage
\bibliographystyle{naturemag}
\bibliography{bib_files/george_covid,bib_files/test_debias,bib_files/equality.bib, bib_files/epimap.bib}       


\end{document}